\documentclass[aps,prl,superscriptaddress,twocolumn,floatfix,tightenlines,showpacs,notitlepage]{revtex4-1}
\usepackage{graphicx}
\usepackage{bm}
\usepackage[normalem]{ulem}
\usepackage{amsfonts}
\usepackage{color}
\usepackage{ulem}
\usepackage{amsmath}    
\usepackage{epsfig}
\usepackage{subfigure}  
\usepackage[breaklinks,colorlinks = true,linkcolor = blue,urlcolor=blue,citecolor=blue]{hyperref}
\usepackage{amssymb}
\usepackage{lineno}
\usepackage{hyperref}
\setcitestyle{square}

\newcommand{\ang}[1]{\ensuremath{\left\langle {#1} \right\rangle}}
\usepackage[dvipsnames]{xcolor}

\newcommand{\icts}{International Centre for
  Theoretical Sciences, Tata Institute of Fundamental Research,
  Bangalore 560089, India}
\newcommand{\mpi}{18/291A, Kottayam 686564, Kerala, India}
\begin{document}
\title{Anomalous diffusion and L\'evy walks distinguish active from inertial turbulence}
\author{Siddhartha Mukherjee}
\email{siddhartha.m@icts.res.in}
\affiliation{\icts}%
\author{Rahul K. Singh}
\email{rksphys@gmail.com}
\affiliation{\icts}%
\author{Martin James}%
\email{martin.james@yahoo.com}
\affiliation{\mpi}%
\author{Samriddhi Sankar Ray}%
\email{samriddhisankarray@gmail.com}
\affiliation{\icts}%

\begin{abstract}

Bacterial swarms display intriguing dynamical states like active turbulence.
Using a hydrodynamic model we now show that such dense active
	suspensions manifest super-diffusion, via L\'evy walks, which
	masquerades as a crossover from ballistic to diffusive scaling in
	measurements of mean-squared-displacements, and is tied to the
	emergence of hitherto undetected oscillatory \textit{streaks} in the flow. Thus,
	while laying the theoretical framework of an emergent advantageous
	strategy in the collective behaviour of microorganisms, our study
	underlines the essential differences between active and inertial
	turbulence.

\end{abstract}

\maketitle

Flowing active matter, resulting from the motility of organisms, cells and particles, forms an intriguing
class of non-equilibrium phenomena
\citep{marchetti2013hydrodynamics,ramaswamy2017active,doostmohammadi2018active}. 
Biological functions like foraging and evasion, that require active agents to
both sample their neighbourhoods and make large jumps to cover ground, become inextricably coupled with collective flow patterns in
dense systems \citep{shellard2020rules}. Active flow driven enhanced diffusion and mixing are also essential for feeding of microorganisms
\citep{humphries2009filter,leptos2009dynamics,lagarde2020colloidal}. Active agents, hence, profit from optimal processes like L\'evy
motion~\citep{humphries2010environmental,volpe2017topography,shlesinger1986levy} characterised 
by long-tailed, self-similar step-size distributions leading to anomalous
diffusion~\citep{klafter2005anomalous} and increased encounter rates
\citep{bartumeus2002optimizing} --- all which can emerge from simple
generative mechanisms~\citep{reynolds2015liberating}---as opposed to
inefficient meandering by random walks limited to classical diffusion. Interestingly, the motion of individual active entities in isolation 
often differ from when they are in large numbers~\citep{knebel2021collective}: Movements of a single 
swimming bacteria can be fundamentally different from its motion in a dense, 
\textit{fluid-like} swarm~\cite{ariel2015swarming}.

A remarkable feature of the collective behaviour of dense active suspensions is the emergence of spatio-temporal structures strongly reminiscent of
inertial 
turbulence~\citep{dombrowski2004self,wensink2012meso,dunkel2013fluid,zhou2014living,wu2017transition,martinez2019selection}: 
Such two-dimensional suspensions are vortical~\citep{giomi2015geometry},
chaotic~\citep{wensink2012meso} with non-Gaussian distributions of velocity
gradients~\citep{giomi2015geometry,james2018turbulence,james2018vortex} and a power-law kinetic energy spectrum~\citep{bratanov2015new,alert2020universal}. These facets of low Reynolds number suspensions have led to a new class of phenomena known as \textit{active
turbulence} \citep{alert2021}.

However, does the similarity between high Reynolds number inertial turbulence and low Reynolds number active turbulence hold even for Lagrangian statistics? This
is a fundamentally important question for two reasons. First, while for
inertial turbulence, Lagrangian (tracer) trajectories, measured through mean
squared displacements (MSD) have a universal
behaviour of purely diffusive self-separation~\citep{xia2013lagrangian}, experiments in active turbulence suggest non-universal signatures of L\'evy
walks and anomalous
diffusion~\citep{wu2000particle,kurtuldu2011enhancement,morozov2014enhanced,ariel2015swarming}. 
Secondly, if such effective biological strategies emerge in active suspensions, why have they remained theoretically undetected and experimentally 
inconclusive so far?

\begin{figure*}
\centering
\includegraphics[width=\linewidth]{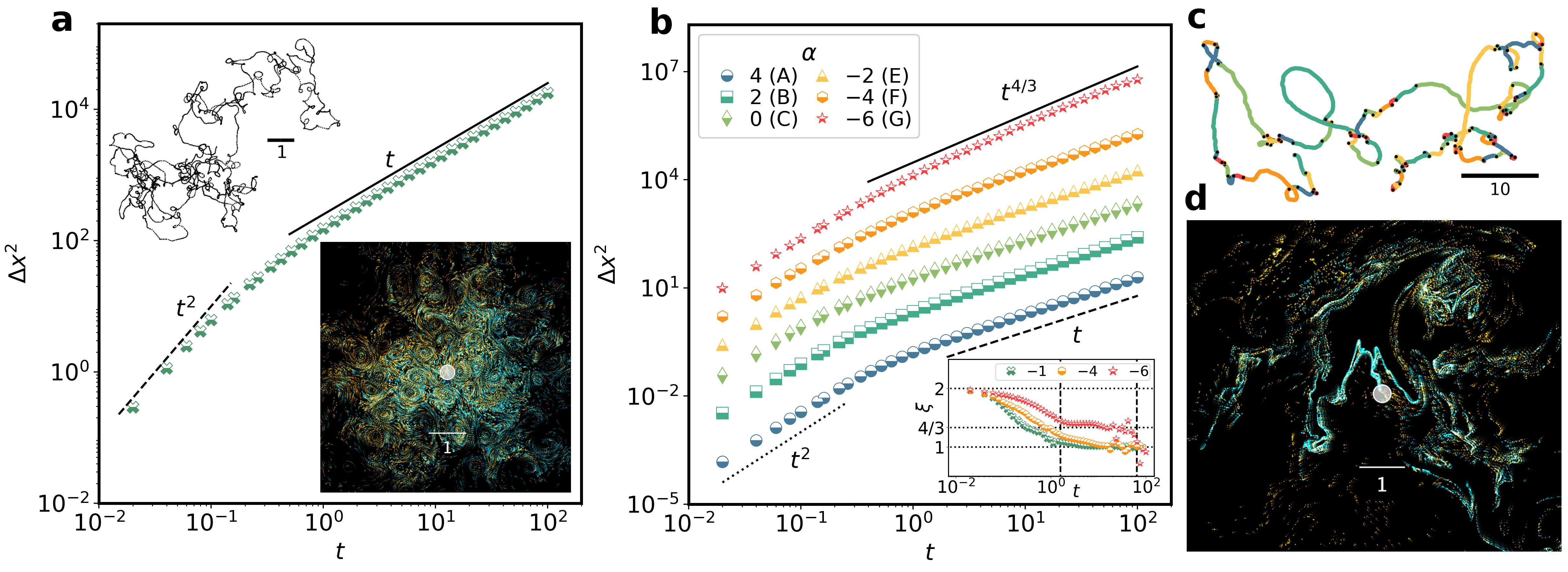}
	\caption{MSD of Lagrangian trajectories (a) for mild activity and (b) 
	with increasing levels of activity (vertically staggered for clarity). \textcolor{black}{Local slopes (b, inset) of $\Delta x^2$: While $\xi$ continuously decreases from $2$ to $1$ for $\alpha\leq -4$, it plateaus (between the vertical lines) at $\xi \approx 4/3$ for $\alpha=-6$.} 
	Representative trajectories for (a) $\alpha = -1$ (upper inset) 
	and (c) $\alpha = -6$ (see text) reflecting the change from diffusive to anomalous behaviour and also seen  
	in snapshots of particle positions (initially localised within the small white disk) at $t \approx 10$ for $\alpha = -1$ (panel (a), lower inset) and (d) $\alpha = -6$; the brightness of the colours reflects the particle density (see movie at \url{https://youtu.be/w3V_cQdvC2k}).}
\label{fig:MSD}
\end{figure*}

Using a continuum hydrodynamic model we now provide definitive answers to
these questions. We show what appears as an \textit{inconsequential}
transition (see Fig.~\ref{fig:MSD}(a)) between the ballistic and diffusive
scaling regimes in MSD measurements is really an
intermediate, anomalous diffusive regime leveraging the crucial biological
advantages of L\'evy walks \citep{kanazawa2020loopy}.  Thus, while low Reynolds number active suspensions may well share
features, at the level of equations and the resultant dynamics, with high
Reynolds number inertial turbulence, active turbulence still allows for emergent behaviour consistent with biological systems striving for efficient searching strategies.  This, we discover, is facilitated by a
synthesis of two basic flow patterns: Novel oscillatory \textit{streaks} responsible for anomalous diffusion and unique to such systems, and vortical features reminiscent of inertial turbulence. 

Dense, active suspensions lend themselves to a generalized hydrodynamic 
description, developed for bacterial swarms \cite{wensink2012meso,dunkel2013minimal}
\begin{equation}
\partial_t {\bf u} + \lambda {\bf u}\cdot \nabla {\bf u}=-\nabla p - \Gamma_0\nabla^2 {\bf u} - \Gamma_2 \nabla^4 {\bf u} - (\alpha + \beta \vert {\bf u} \vert^2) {\bf u} \label{eq:genHyd}
\end{equation}
where the incompressible velocity field ${\bf u}({\bf x},t)$ (with $\nabla\cdot\bf u=0$), is a coarse-grained description of the motility
of dense, active (bacterial) suspensions. Here, $\lambda>0$ corresponds to
pusher-type bacteria and the $\Gamma-$terms are responsible for quasi-chaotic
pattern formation via stress-induced instabilities (when $\Gamma_0,\Gamma_2>0$)
in the bacterial
system~\citep{swift1977hydrodynamic,wensink2012meso,simha2002hydrodynamic,linkmann2020condensate,bratanov2015new}.
The last term adds Toner-Tu~drive~\citep{TT95,TT98}, where $\beta$ needs to be
positive for stability, while the activity $\alpha$ can take both positive
(Ekman friction) and negative (active injection) values.
In our direct numerical simulations (see Supplemental Material), $\Gamma_0$ and $\Gamma_2$ are fixed according to experimental length and time scales~\cite{wensink2012meso}, while other parameters are varied to
obtain the flow fields explored in this study. The characteristic
length and time scales associated with the linear instability in
\eqref{eq:genHyd} are $L_\Gamma =\sqrt{{\Gamma_2}/{\Gamma_0}}$ and $\tau_\Gamma =
{\Gamma_2}/{\Gamma_0^2}$~\cite{wensink2012meso}, while the activity length and
time scales are given by  $1/\sqrt{\vert\alpha\vert \beta}$ and $1/|\alpha|$, 
respectively. In conformity with the previous studies using this model, we
present our results in simulation units.

Figure~\ref{fig:MSD}(a) shows the MSD $\Delta x^2 = \ang{\vert\vert\mathbf{x}(t) -
\mathbf{x}(0)\vert\vert^2}$ (where $\vert\vert\cdot\vert\vert$ and
$\ang{\cdot}$ denote the Euclidean norm and ensemble averaging over all
particles, respectively) for an active suspension with $\alpha = -1.0$. This
scaling behaviour seems consistent with inertial turbulence: A ballistic
regime $\Delta x^2 \sim t^2$ crossing over to a diffusive regime $\Delta x^2
\sim t$~\citep{joy2020friction}. Individual trajectories 
(upper inset, Fig.~\ref{fig:MSD}(a)), similarly, show diffusive meandering.

However, some  experiments~\citep{wu2000particle,ariel2015swarming} on dense bacterial swarms
provide strong evidence of anomalous diffusion $\Delta x^2 \sim t^\xi$, with $1
< \xi < 2$. This raises the question whether what is seen as a crossover from
ballistic to diffusive behaviour  is
\textit{actually} masking an intermediate, anomalous diffusive regime. 

To uncover the true behaviour of such suspensions, we perform several
simulations on domain sizes $20 \leq L \leq 80$ with $-6 \leq \alpha
\leq 4$, seeded with tracers that evolve as
$\mathrm{d}\mathbf{x}(t)/\mathrm{d}t = \mathbf{u}(\mathbf{x},t)$, and measure
the associated MSD as shown in Fig.~\ref{fig:MSD}(b).  For $\alpha \geq 0$,
corresponding to an Ekman friction effect, and for modest activity $0 < \alpha \leq -2$  (Cases A -- E), we see little evidence of anomalous
diffusion. However, as we increase the activity further (Case F), the first
signatures of an intermediate regime appear. This observation is validated for
Case G where the $\Delta x^2$ shows a convincing super-diffusive regime, and
local slope analysis (Fig.~\ref{fig:MSD}(b), inset) gives $\xi = \frac{d {\rm
Log} \Delta x^2}{d {\rm Log} t} =  1.31 \pm 0.08 \approx 4/3$, for close to two
decades before giving way to diffusion. This scaling, we recall, is not
inconsistent with recent experimental
measurements~\citep{wu2000particle,ariel2015swarming} and simplified model predictions~\citep{ariel2017chaotic,ariel2020conservative} which suggest similar super-diffusion for dense suspensions of motile bacteria. It is important to alert the reader that whether these are indeed non-trivial fixed points in the renormalisation 
group sense would require an exponent flow~\citep{panja2015complex} or more sophisticated data analysis through 
asymptotic extrapolation~\cite{hoeven,pauls-frisch,entire,nelkin}; given the high precision data required for such 
approaches to be beyond speculative, we refrain from this analysis here.

The emergence of anomalous diffusion, as suspensions
become more active, ought to carry its signature in Lagrangian trajectories, providing a crucial link 
in understanding how bacterial colonies forage and avoid hostile environments.
In Fig.~\ref{fig:MSD}(c) we show a representative
trajectory of a tracer corresponding to a highly active suspension with $\alpha = -6$ which, in sharp contrast to trajectories in mildly active suspensions (Fig.~\ref{fig:MSD}(a), upper inset), shows short diffusive behaviour punctuated by long ``steps'' indicative of anomalous diffusion. This is clearly illustrated in the movies of trajectories for different activities (see \url{https://youtu.be/G97ndUVeNRQ}).

Figure~\ref{fig:MSD}(c) shows a trajectory divided into segments (shown in different colours), that are identified every time the trajectory turns (points marked in black) by an angle $\theta$ greater than a threshold angle
$\theta_c$; we choose  $\theta_c = 30^\circ$ for illustration in Fig.~\ref{fig:MSD}(c).  
While trajectories evolve mostly with small variations in $\theta$ for mild suspensions, 
increasing activity makes sharper turns more probable (see Supplemental Material).

The changing nature of particle trajectories and transition from normal to anomalous diffusion is also reflected in the spreading of an initially localised \textit{puff} of particles (see \url{https://youtu.be/w3V_cQdvC2k}). In Fig.~\ref{fig:MSD}(d) we show a
snapshot of 20,000 particles, initially localised within the small white disk at the center,
at time $t \approx 10$ for Case G ($\alpha = -6$); the brightness of the colours is a measure of the local density of
particles. The evolution of the puff in highly active suspensions is far more ``irregular'' than what is seen, under identical conditions, at low activity ($\alpha = -1$, Fig.~\ref{fig:MSD}(a), lower inset) where they spread through diffusion.

\begin{figure}
\centering
	\includegraphics[width=\linewidth]{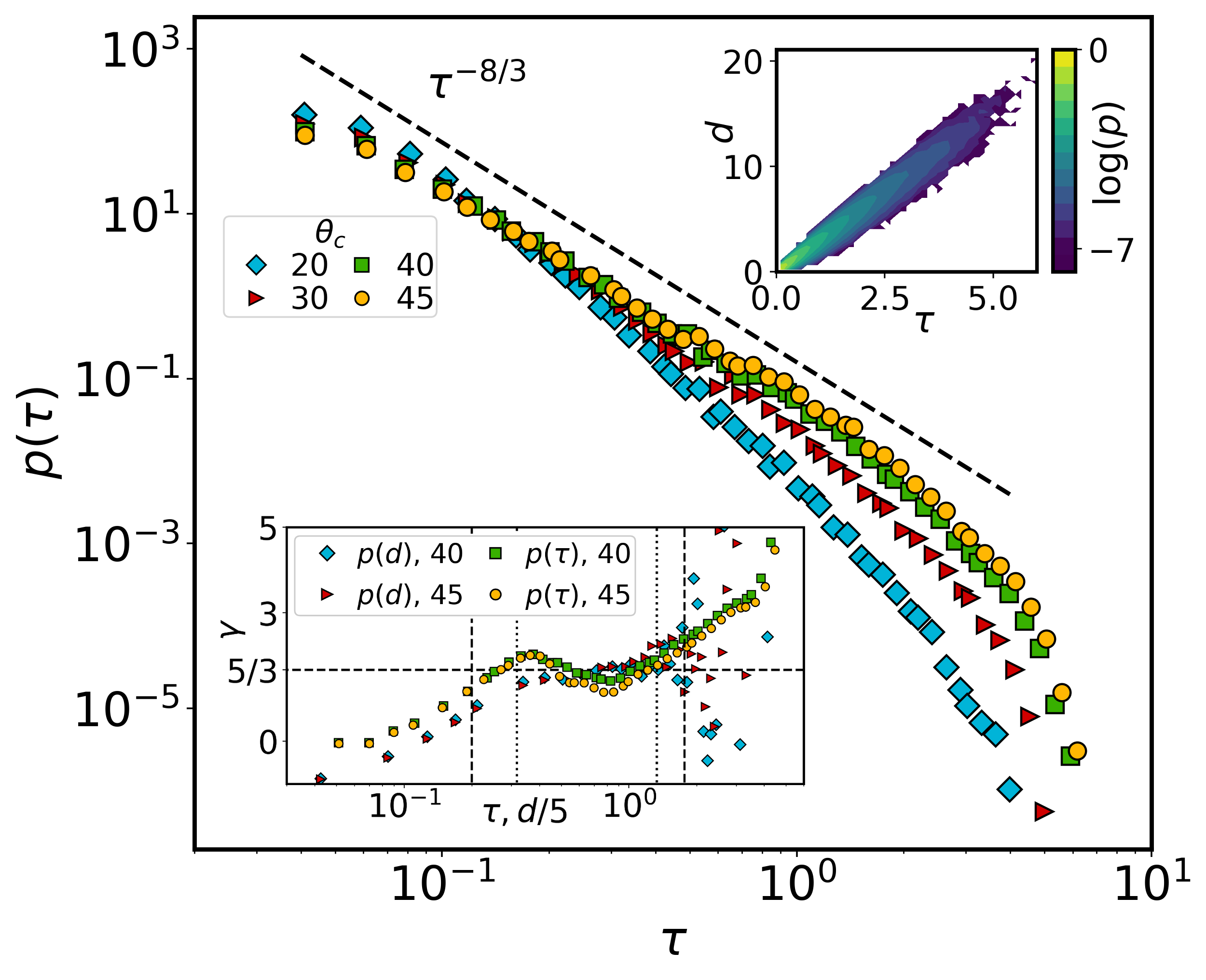}
	\caption{For highly active suspensions (Case G), probability distributions $p(\tau)$ of the waiting time $\tau$ 
	with a $\tau^{-8/3}$ scaling (dashed line), as a guide to the eye, for different $\theta_c$. (Lower inset) Local slope analysis of $p(\tau)$ and $p(d)$ (x-axis rescaled for visualisation) yields $\gamma = 1.7 \pm 0.3$ and $\gamma = 1.6 \pm 0.2$, respectively, consistent with L\'evy walk estimates, showing a scaling range (between dashed ($p(\tau)$) and dotted ($p(d)$) vertical lines) for about a decade.
	(Upper inset) Joint probability distribution of $d$ and $\tau$ confirming a finite velocity through	the near linear relationship between the two.}
\label{fig:levy}
\end{figure}

This inevitably leads to the question whether these visual cues are
stemming from \textit{real} L\'evy walks, marked by power-law
step size and equivalently, because of an approximately constant velocity,
waiting time distributions~\cite{shlesinger1987levy}. A convincing argument in favour of L\'evy walks is the distribution $p(\tau)$ of waiting times $\tau$, which ought to show a significant range of scaling of the form
$p(\tau) \sim \tau^{-\gamma-1}$ for a reasonably large spread
in the choice of the threshold angle $\theta_c$ which determines a ``turn''.
Fig.~\ref{fig:levy} confirms the existence of such a power-law and
a local slope analysis $\gamma = -1 - \frac{d {\rm Log}p(\tau)}{d {\rm Log}
\tau}$ (Fig.~\ref{fig:levy}, lower inset) shows scaling for about a decade with $\gamma = 1.7 \pm 0.3$. Similarly, the probability distribution $p(d)$ of step sizes, for the more active suspensions, 
follow a scaling $p(d) \sim d^{-\gamma - 1}$. Hence, in the same inset, we also
show local slopes obtained from $p(d)$ (with the x-axis rescaled to \textcolor{black}{$d/5$} for ease of comparison), showing a comparable extent of
scaling, from which we obtain $\gamma = 1.6 \pm 0.2$.  Importantly, this scaling exponent $\gamma \approx 5/3$  when coupled with the anomalous MSD exponent $\xi \approx 4/3$, satisfies the L\'evy walk constraint $\gamma +\xi = 3$~\citep{zaburdaev2015levy}. Finally, the joint distribution of flight lengths $d$ and waiting times $\tau$ between turns in the trajectories for active suspensions (Fig.~\ref{fig:levy},
upper inset) shows an almost linear scaling reflecting a constant system velocity.  All of these are clear, unambiguous indicators of L\'evy walks~\cite{footnote1}. 

\begin{figure}
\centering
\includegraphics[width=\linewidth]{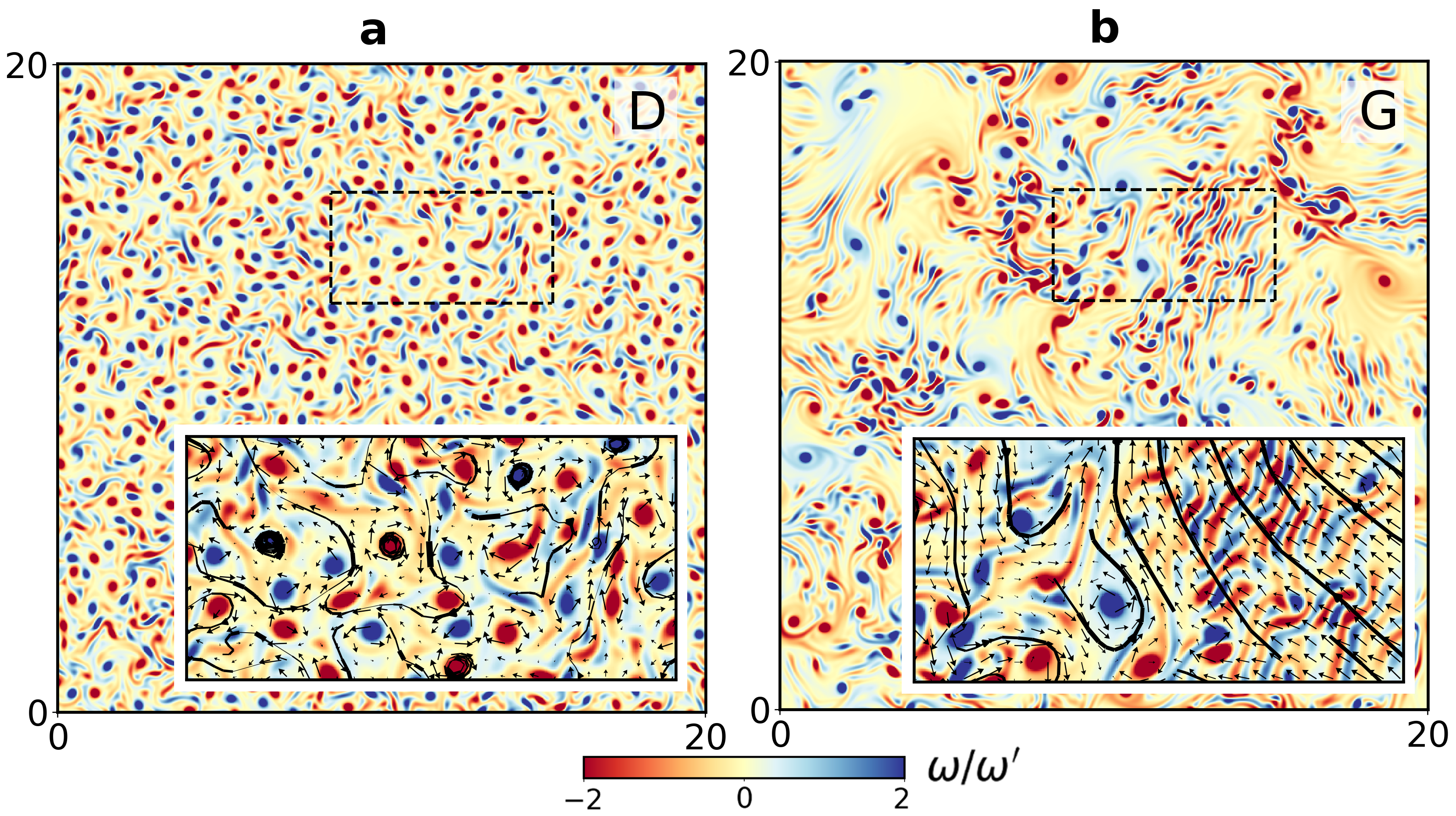}
\caption{Representative snapshots of vorticity fields (and their magnified
	sections with velocity vectors as arrows and instantaneous streamlines as solid lines) for (a) Case D ($\alpha = -1$),
	and (b) Case G ($\alpha = -6$). In the latter, highly active suspensions
	a new feature shows up: The vorticity field is now populated by oscillatory
	\textit{streaks} (see also Fig.~\ref{fig:extreme} and \url{https://youtu.be/gqhT-xf_tAc}).} 
	\label{fig:omega}
\end{figure}

Does the emergence of (Lagrangian) anomalous scaling carry tell-tale signs in the vorticity field
${\boldsymbol \omega} = \nabla \times {\bf u}$? Figure~\ref{fig:omega} (see \url{https://youtu.be/gqhT-xf_tAc} for movies of evolving fields) shows slices of the vorticity field, for Cases
D and G. While the vorticity fields appear, overall,
\textit{disorganized}---active turbulence---a closer inspection reveals a
pattern hitherto undetected. With increasing activity, the dense
packing of diffused vortices, which mainly appear as \textit{spots},
gives way to \textit{sharper} spots with trailing wisps of vorticity
\textit{streaks} (Case D). For highly active suspensions (Case G), there is an
equal preponderance of spots and streaks. Curiously, these streaks appear with
alternating signs in a periodic fashion with a characteristic length scale
$\delta$.

The probability distribution of vorticity $p(\omega)$ is non-Gaussian (similarly to inertial turbulence) with fat tails which fall-off
as a power-law, becoming more pronounced with increasing activity (see Supplemental Material). This analogy with inertial turbulence extends to 
the energy spectrum scaling as $k^{-5/3}$ scaling at high activity~\cite{Pandit2009,pandit2017overview} (see Supplemental Material).

\begin{figure} 
	\includegraphics[width=\linewidth]{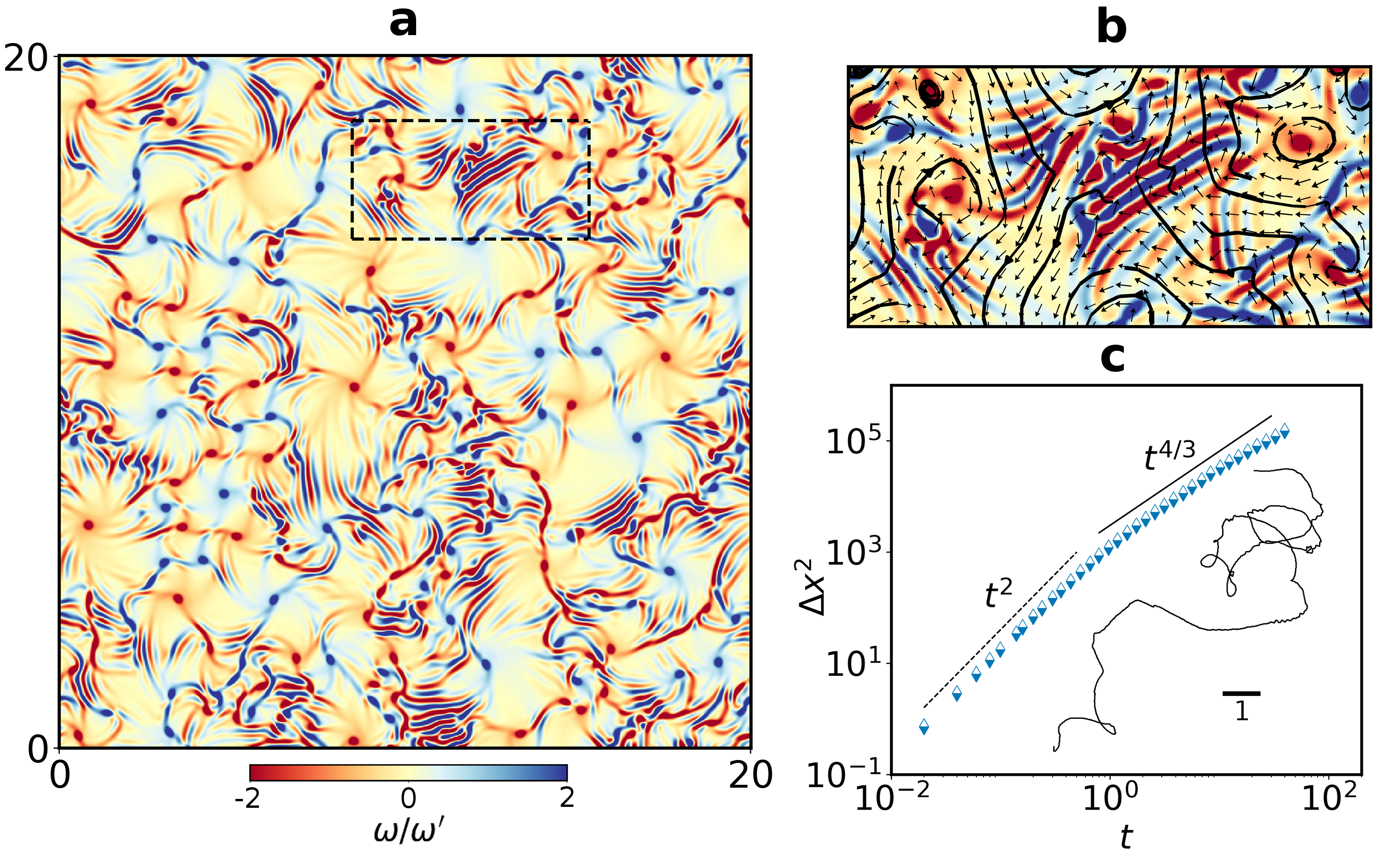}
	\caption{Extremely active suspensions shows that (a) the vorticity field is
	dominated by \textit{streaks} interspersed with fewer vortical
	\textit{spots} (magnified view in panel (b) with velocity vectors as arrows and instantaneous streamlines as solid lines) and is associated with (c) robust anomalous diffusion $\Delta x^2 \sim t^{4/3}$, reflected in a representative trajectory (inset).} 
	\label{fig:extreme}
\end{figure}

Is there a causal link between the alternating \textit{streaks} and the
anomalous diffusion that we observe? Careful measurements certainly suggest so.
While a definitive answer is beyond the scope of this work, 
the instantaneous, jet-like velocity and elongated streamlines, orthogonal 
to the streaks,  (Fig.~\ref{fig:omega} insets) would in the absence of dynamics ensure the persistence of trajectories, often over lengths $\mathcal{O}(L/2)$ (consistent with the scaling extent in $p(d)$), while the vortical patches serve to reorient 
trajectories. Na\"ively, this suggests streak regions as origins of anomalous diffusion, while their absence (low activity) leads to 
classical diffusion. Importantly, these streaks are distinct from the ``laning'' phenomena in some active suspensions~\citep{wensink2012emergent}. In order to further
test the role of streaks, we push the system to an \textit{extreme} case (with
$\alpha=-16.66$ and $\beta=22.22$), which shows that the emergence of more
distinct, pattern forming streaks (Fig.~\ref{fig:extreme}(a), with a magnified
view in panel (b)) is accompanied by a robust super-diffusive $\Delta x^2 \sim
t^{4/3}$ regime (Fig.~\ref{fig:extreme}(c)). Unsurprisingly, the trajectories
reflect this anomalous diffusion through clear L\'evy walks
(see inset in Fig.~\ref{fig:extreme}(c)), and are also found to have power-law
distributions of $d$ and $\tau$. Thus, we show that active systems are actually
super-diffusive and what may be mistaken as a cross-over from ballistic to
diffusive behaviour masks the most important and non-trivial aspects of such
systems.

While the origin of the oscillatory streaks remains to be established, it seems
reasonable to assume that the length scale $\delta$, determining the
alternating pattern of streaks, is influenced by the activity which sets the
characteristic velocity for global polar ordering $v_0 = \sqrt{\vert \alpha
\vert/\beta}$~\citep{wensink2012meso}. Since the dominant time-scale is set by
activity,  $\delta \simeq 1/\sqrt{\vert\alpha\vert \beta}$. Given the heuristic
nature of this argument, we made careful measurements of the ``wavelength'' of
the oscillations for a range of parameters and found that they are in
reasonable agreement with our conjectured estimate. Why the system senses this
length scale will perhaps be found when the origins of these oscillatory
streaks are systematically known, and the question of universality (or not) of
such oscillatory patterns for different classes of active systems is an
important one. While the specific model~\cite{toner} studied in this work seems
to allow for an instability that triggers spatio-temporal chaotic states with
bands of opposite polarities which show up as spots and
streaks~\cite{PC-Maitra}, the precise mechanisms involved are left for future
work.

Nature exploits L\'evy movements and anomalous diffusion, across scales, ranging from the
microscopic to ecological~\citep{reynolds2018current}, and across taxa, from
systems comprising individual agents like midge
swarms~\cite{reynolds2016swarm}, migrating metastatic cancer
cells~\citep{huda2018levy}, living cancer cells~\cite{gal2010} and intracellular DNA transport~\citep{muralidharan2021intracellular}, foraging marine predators~\citep{sims2008scaling}
and expanding colonies of seemingly immobile beach
grasses~\citep{reijers2019levy}, to dense systems with collective flow states
like swimming bacteria~\citep{ariel2015swarming}. Yet, detecting L\'evy walks
theoretically in active turbulence has remained elusive, despite some
experimental results strongly suggestive of their existence. While recent work
using a particle-based active model~\citep{kanazawa2020loopy} reconciles some
of these findings, the lack of consistency with the most general hydrodynamic
framework to describe active suspensions is surprising. We uncover why this is,
showing that while it is true that anomalous diffusion is hardly detectable
(though incipient in the light of our results) for mildly active suspensions,
such systems exhibit distinct  L\'evy walks and super-diffusion when nudged to
higher levels of activity.

Our observations of the vorticity spots and streaks, a basis of future theories of emergent anomalous diffusion, 
provide a template for clearly identifiable
structures experimentally. Thus experiments would be able to probe how
universal these patterns are, aiding a more robust understanding of why they
emerge and whether they are central to the super-diffusive behaviour of such
systems. Recent works, which derive the
hydrodynamic model presented here from microswimmer dynamics, show that the activity is
related, among other factors, to their individual motility~\cite{heidenreich2016hydrodynamic,reinken2018derivation}.
Thus high activity might be achieved by tuning the motility.
A potential candidate for an experiment is a spermatozoa suspension~\cite{james2020emergence}, which also exhibit a turbulent phase~\cite{creppy2015turbulence} and whose motility can be controlled by changing the ambient temperature~\cite{alavi2005sperm}.

Finally, we underline what sets active turbulence apart from inertial fluid
turbulence and thus the limitations of drawing equivalences between the two.
The Lagrangian picture, arising from our work, marks a crucial departure in
the analogy in a rather counter-intuitive way. This is because for 
Eulerian statistics, increasing activity results in a \textit{more
intermittent} vorticity field and accompanying
power-laws of energy spectra (see Supplemental Material) like in
two-dimensional fluid turbulence.
However, this  \textit{``increasingly turbulent
state''} in active systems is paradoxically accompanied by \textit{persistent}
super-diffusion, L\'evy walks and structural changes in the vorticity field
which have no known counterparts in inertial turbulence. It would be interesting, however, to see how this distinction manifests itself in other 
measurements such as pair-particle dispersion~\citep{boffetta2000pair}.

Before concluding, we recall that our MSD exponent $\xi \approx 4/3$ (at high
activity) is consistent with those predicted for 1D Hamiltonian
systems~\cite{spohn}. Whether this suggests a possible underlying 
universality is unclear. 
Recent theoretical
studies of the walk time distribution~\cite{MironPRL2020,MironPRR2020} show that local dynamics
are central to super-diffusion with $\gamma = 3/2$. 
While we detect $\gamma \approx 5/3$, it
is possible that this exponent is one of an intermediate asymptotics (in
$\alpha$) which may converge to $3/2$ for sufficiently high activity and 
diffusive behaviour appears as a
pre-asymptotic correction to the asymptotic super-diffusive regime~\cite{PC-Miron} that we report.

Our work suggests that activity is geared for manifesting optimality: Yet
another way biological systems continue to defy bounds on inanimate matter, now
in a turbulence-like flow state. This gives new direction to the assessment of
what is truly turbulent, and universal, in low Reynolds number active flows.

\paragraph*{Acknowledgements:} We thank G. Ariel, E. Barkai, A. Kundu, and D. Mukamel for discussions, and especially, A. Maitra
and A. Miron for several suggestions and comments on our
work. The simulations were performed on the ICTS clusters {\it Tetris}, and {\it Contra} as well as the work stations from the project ECR/2015/000361: {\it
Goopy} and {\it Bagha}. SSR acknowledges SERB-DST (India) project DST (India)
project MTR/2019/001553 for financial support.  The authors acknowledges the
support of the DAE, Govt. of India, under project no.  12-R\&D-TFR-5.10-1100
and project no.  RTI4001

\section*{Supplemental Material}
\textbf{Direct Numerical Simulations}\\

The generalized hydrodynamic model, is solved using a pseudo-spectral
method~\citep{james2018vortex}.  Simulations are performed on a periodic domain
of sizes $1024^2$ and $4096^3$ grid points, with a physical extent of $L=20$ and $L = 80$, respectively, for a duration
of $5\times 10^5$ iterations, with a time-step of $\delta t = 0.0002$. To be
consistent with previous work
\citep{wensink2012meso,james2018vortex,joy2020friction}, we set $\Gamma_0 =
0.045$, $\Gamma_2 = \Gamma_0^3$, $\beta = 0.5$, $\lambda = 3.5$. Cases A to G
correspond to $\alpha \in \left\lbrace 4, 2, 0, -1, -2, -4, -6\right\rbrace$. 
We additionally simulate Case Z, after a parameter search, with $\beta=22.22$
and $\alpha = -16.66$, which is to exemplify an extreme state dominated by
streaks, leading to robust anomalous diffusion. 

After a spinup period of $2\times10^4$ iterations, the flow is seeded with
$10^5$ tracer particles randomly distributed in the domain, which follow the
equation \textcolor{black}{$\mathrm{d}\mathbf{x}(t)/\mathrm{d}t = \mathbf{u}(\mathbf{x},t)$}~\cite{perl2011,ssr2011},
integrated with a 4th-order Runge-Kutta scheme. Field quantities are
interpolated to off-grid locations using bilinear interpolation, and data is
stored every $100$ iterations to report well-converged statistics.\\

\textbf{Quantifying Trajectories}\\

Each trajectory is divided into individual walking segments (as illustrated in the main text) that are identified by a threshold on 
the turning angle $\theta$ at each time $t$. The angles $\theta(t)$ are calculated along trajectories as 
\begin{equation}
\cos\left(\theta(t)\right) = \frac{ \Delta \mathbf{r}(t)\cdot \Delta \mathbf{r}(t+\Delta t) }{\vert\Delta \mathbf{r}(t)\vert \vert\Delta \mathbf{r}(t+\Delta t)\vert}
\end{equation}
where $\Delta \mathbf{r}(t^\prime)$ is the displacement vector connecting points $\mathbf{r}(t^\prime - \Delta t)$ and $\mathbf{r}(t^\prime)$, and $\Delta t$ is the temporal coarsening, and results were found similar for $\Delta t = 2\delta t , 3\delta t$ and $4\delta t$. Individual segments and turns can be identified using a simple threshold of $\theta(t)>\theta_c$, although the exact segments are rather sensitive to the choice of $\theta_c$. A filtering is done to identify clusters of two or three successive points that are all identified as ``turns'' (usually occurs during sharp turns), where only the point with the highest turning angle is retained. The waiting time $\tau$ is calculated as the time between successive turns.\\

\textbf{Supplemental Figures}

\begin{figure}[!h]
\centering
\includegraphics[width=0.85\linewidth]{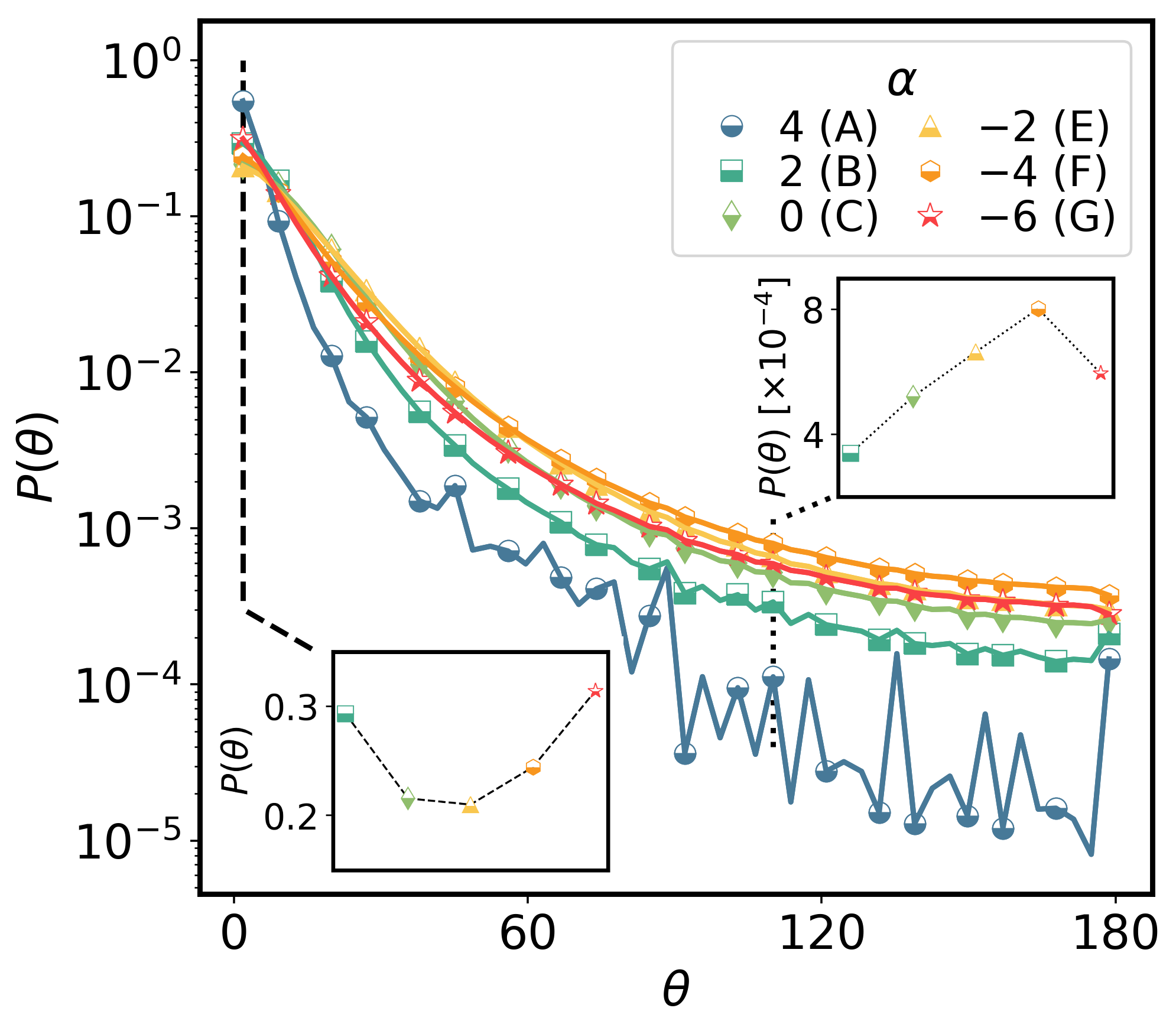}
	\caption{\textbf{Distribution of turning angles.} The probability distributions $P(\theta)$ along trajectories show that activity increases the probability of sharper turns. Moreover, the propensity for small changes in $\theta$ (Case A) first decreases with activity (up to Case E), and then increases again (Cases F and G), which, together with the prevalence of sharp turns at higher activity, reflects L\'evy walk-like behaviour.}
\label{fig:angles}
\end{figure}

\begin{figure}
\centering
\includegraphics[width=\linewidth]{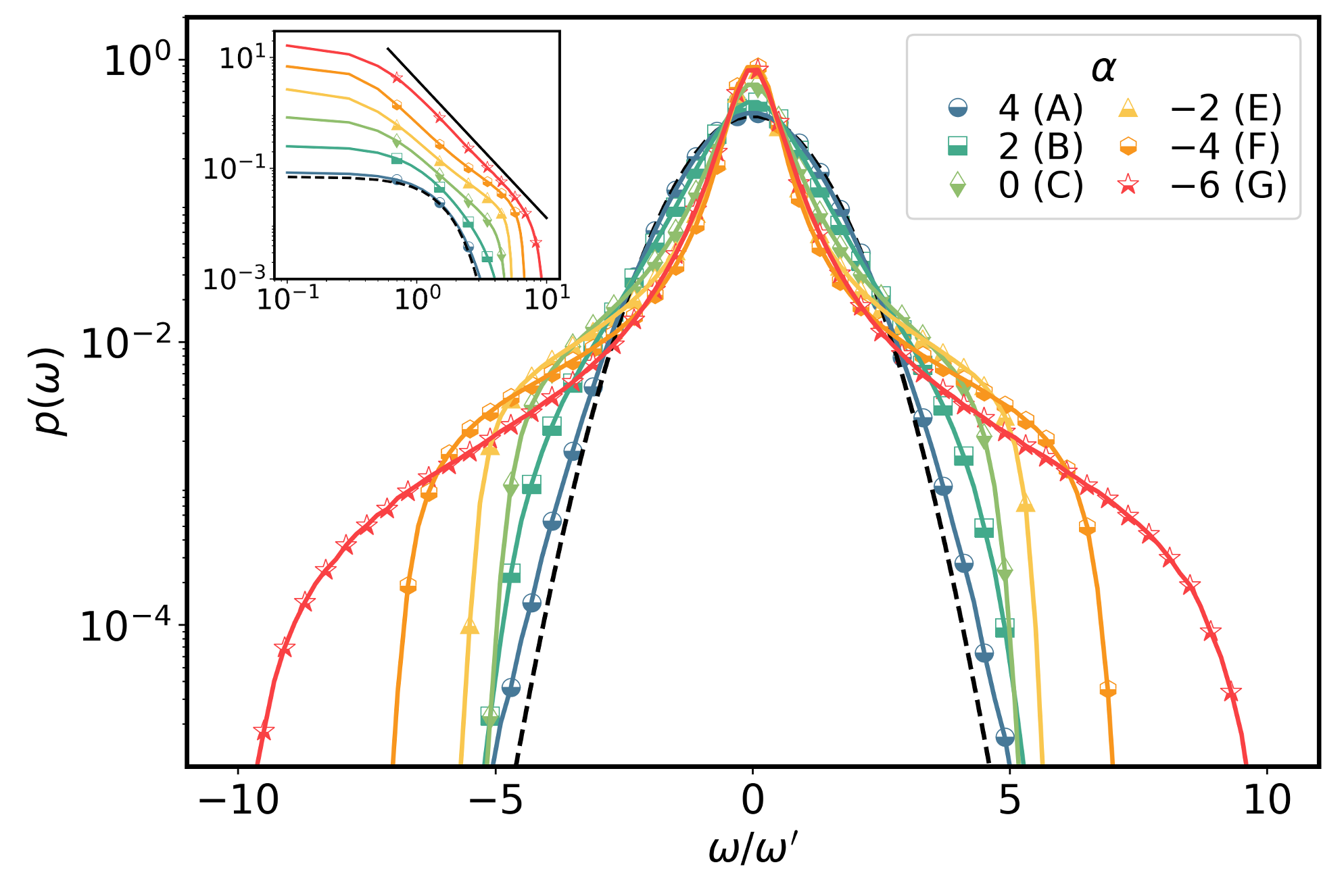}
	\caption{\textbf{Vorticity distribution.} The statistics of the $\boldsymbol {\omega}$ field is quantified by the probability density function $p(\omega)$ and it shows non-Gaussian, fat tails with power-laws (see inset, where the curves have been vertically staggered for clarity) as suspensions become more active. This increasing intermittency with higher activity is analogous to the effect of increasing Reynolds number in inertial turbulence.}
\label{fig:vort}
\end{figure}

\begin{figure}
\centering
\includegraphics[width=\linewidth]{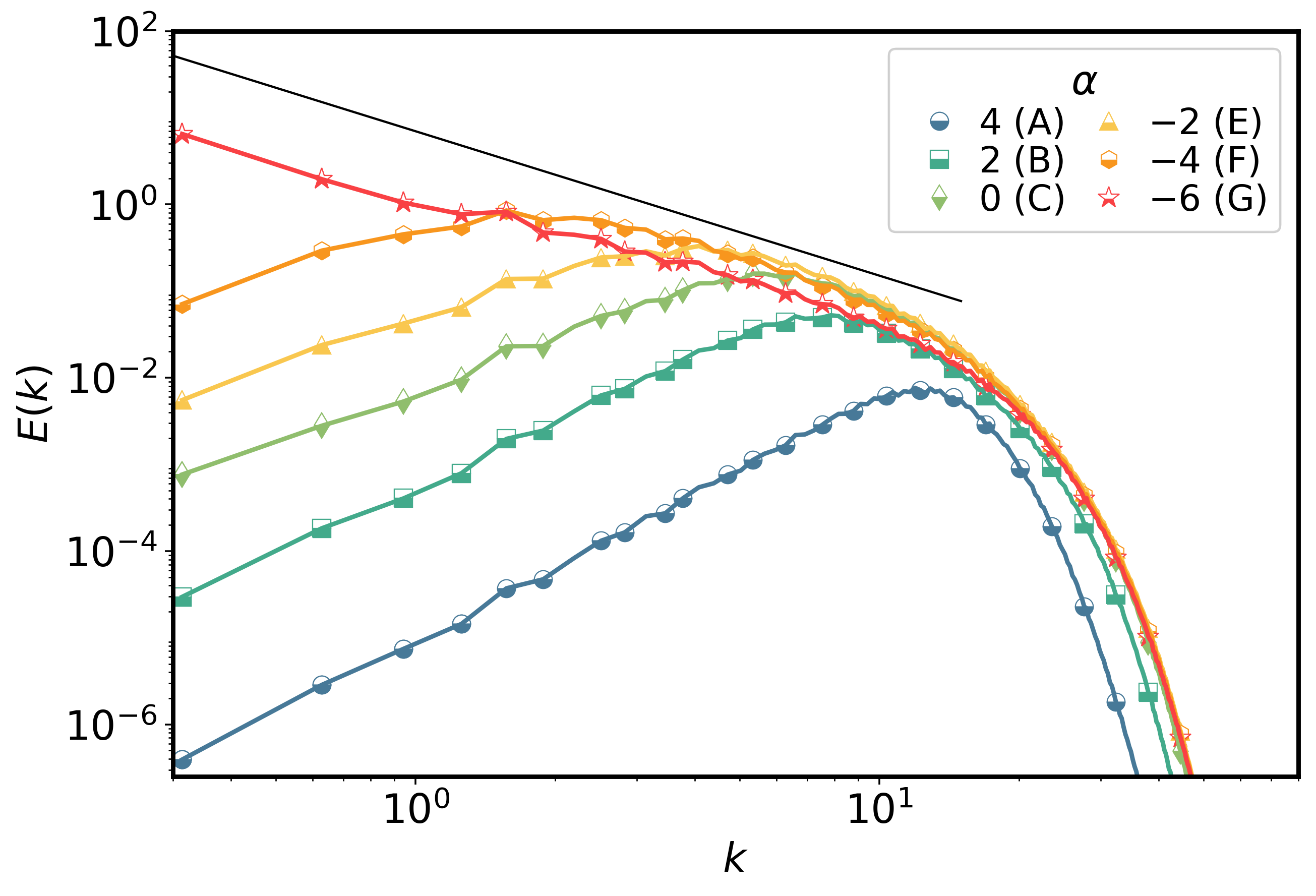}
	\caption{\textbf{Energy spectra.} The kinetic energy spectrum $E(k)$ asymptotically approaches a $k^{-5/3}$ power-law as the level of activity becomes high, which is again analogous to the energy spectrum in high Reynolds number inertial turbulence with an inverse-cascade.}
\label{fig:kespectra}
\end{figure}

\bibliography{references}

\begin{thebibliography}{75}%
\makeatletter
\providecommand \@ifxundefined [1]{%
 \@ifx{#1\undefined}
}%
\providecommand \@ifnum [1]{%
 \ifnum #1\expandafter \@firstoftwo
 \else \expandafter \@secondoftwo
 \fi
}%
\providecommand \@ifx [1]{%
 \ifx #1\expandafter \@firstoftwo
 \else \expandafter \@secondoftwo
 \fi
}%
\providecommand \natexlab [1]{#1}%
\providecommand \enquote  [1]{``#1''}%
\providecommand \bibnamefont  [1]{#1}%
\providecommand \bibfnamefont [1]{#1}%
\providecommand \citenamefont [1]{#1}%
\providecommand \href@noop [0]{\@secondoftwo}%
\providecommand \href [0]{\begingroup \@sanitize@url \@href}%
\providecommand \@href[1]{\@@startlink{#1}\@@href}%
\providecommand \@@href[1]{\endgroup#1\@@endlink}%
\providecommand \@sanitize@url [0]{\catcode `\\12\catcode `\$12\catcode
  `\&12\catcode `\#12\catcode `\^12\catcode `\_12\catcode `\%12\relax}%
\providecommand \@@startlink[1]{}%
\providecommand \@@endlink[0]{}%
\providecommand \url  [0]{\begingroup\@sanitize@url \@url }%
\providecommand \@url [1]{\endgroup\@href {#1}{\urlprefix }}%
\providecommand \urlprefix  [0]{URL }%
\providecommand \Eprint [0]{\href }%
\providecommand \doibase [0]{http://dx.doi.org/}%
\providecommand \selectlanguage [0]{\@gobble}%
\providecommand \bibinfo  [0]{\@secondoftwo}%
\providecommand \bibfield  [0]{\@secondoftwo}%
\providecommand \translation [1]{[#1]}%
\providecommand \BibitemOpen [0]{}%
\providecommand \bibitemStop [0]{}%
\providecommand \bibitemNoStop [0]{.\EOS\space}%
\providecommand \EOS [0]{\spacefactor3000\relax}%
\providecommand \BibitemShut  [1]{\csname bibitem#1\endcsname}%
\let\auto@bib@innerbib\@empty
\bibitem [{\citenamefont {Marchetti}\ \emph {et~al.}(2013)\citenamefont
  {Marchetti}, \citenamefont {Joanny}, \citenamefont {Ramaswamy}, \citenamefont
  {Liverpool}, \citenamefont {Prost}, \citenamefont {Rao},\ and\ \citenamefont
  {Simha}}]{marchetti2013hydrodynamics}%
  \BibitemOpen
  \bibfield  {author} {\bibinfo {author} {\bibfnamefont {M.~C.}\ \bibnamefont
  {Marchetti}}, \bibinfo {author} {\bibfnamefont {J.-F.}\ \bibnamefont
  {Joanny}}, \bibinfo {author} {\bibfnamefont {S.}~\bibnamefont {Ramaswamy}},
  \bibinfo {author} {\bibfnamefont {T.~B.}\ \bibnamefont {Liverpool}}, \bibinfo
  {author} {\bibfnamefont {J.}~\bibnamefont {Prost}}, \bibinfo {author}
  {\bibfnamefont {M.}~\bibnamefont {Rao}}, \ and\ \bibinfo {author}
  {\bibfnamefont {R.~A.}\ \bibnamefont {Simha}},\ }\href {\doibase
  10.1103/RevModPhys.85.1143} {\bibfield  {journal} {\bibinfo  {journal}
  {Reviews of Modern Physics}\ }\textbf {\bibinfo {volume} {85}},\ \bibinfo
  {pages} {1143} (\bibinfo {year} {2013})}\BibitemShut {NoStop}%
\bibitem [{\citenamefont {Ramaswamy}(2017)}]{ramaswamy2017active}%
  \BibitemOpen
  \bibfield  {author} {\bibinfo {author} {\bibfnamefont {S.}~\bibnamefont
  {Ramaswamy}},\ }\href {\doibase 10.1088/1742-5468/aa6bc5} {\bibfield
  {journal} {\bibinfo  {journal} {Journal of Statistical Mechanics: Theory and
  Experiment}\ }\textbf {\bibinfo {volume} {2017}},\ \bibinfo {pages} {054002}
  (\bibinfo {year} {2017})}\BibitemShut {NoStop}%
\bibitem [{\citenamefont {Doostmohammadi}\ \emph {et~al.}(2018)\citenamefont
  {Doostmohammadi}, \citenamefont {Ign{\'e}s-Mullol}, \citenamefont {Yeomans},\
  and\ \citenamefont {Sagu{\'e}s}}]{doostmohammadi2018active}%
  \BibitemOpen
  \bibfield  {author} {\bibinfo {author} {\bibfnamefont {A.}~\bibnamefont
  {Doostmohammadi}}, \bibinfo {author} {\bibfnamefont {J.}~\bibnamefont
  {Ign{\'e}s-Mullol}}, \bibinfo {author} {\bibfnamefont {J.~M.}\ \bibnamefont
  {Yeomans}}, \ and\ \bibinfo {author} {\bibfnamefont {F.}~\bibnamefont
  {Sagu{\'e}s}},\ }\href {\doibase 10.1038/s41467-018-05666-8} {\bibfield
  {journal} {\bibinfo  {journal} {Nature communications}\ }\textbf {\bibinfo
  {volume} {9}},\ \bibinfo {pages} {1} (\bibinfo {year} {2018})}\BibitemShut
  {NoStop}%
\bibitem [{\citenamefont {Shellard}\ and\ \citenamefont
  {Mayor}(2020)}]{shellard2020rules}%
  \BibitemOpen
  \bibfield  {author} {\bibinfo {author} {\bibfnamefont {A.}~\bibnamefont
  {Shellard}}\ and\ \bibinfo {author} {\bibfnamefont {R.}~\bibnamefont
  {Mayor}},\ }\href {\doibase 10.1098/rstb.2019.0387} {\bibfield  {journal}
  {\bibinfo  {journal} {Philosophical Transactions of the Royal Society B}\
  }\textbf {\bibinfo {volume} {375}},\ \bibinfo {pages} {20190387} (\bibinfo
  {year} {2020})}\BibitemShut {NoStop}%
\bibitem [{\citenamefont {Humphries}(2009)}]{humphries2009filter}%
  \BibitemOpen
  \bibfield  {author} {\bibinfo {author} {\bibfnamefont {S.}~\bibnamefont
  {Humphries}},\ }\href {\doibase 10.1073/pnas.0809063106} {\bibfield
  {journal} {\bibinfo  {journal} {Proceedings of the National Academy of
  Sciences}\ }\textbf {\bibinfo {volume} {106}},\ \bibinfo {pages} {7882}
  (\bibinfo {year} {2009})}\BibitemShut {NoStop}%
\bibitem [{\citenamefont {Leptos}\ \emph {et~al.}(2009)\citenamefont {Leptos},
  \citenamefont {Guasto}, \citenamefont {Gollub}, \citenamefont {Pesci},\ and\
  \citenamefont {Goldstein}}]{leptos2009dynamics}%
  \BibitemOpen
  \bibfield  {author} {\bibinfo {author} {\bibfnamefont {K.~C.}\ \bibnamefont
  {Leptos}}, \bibinfo {author} {\bibfnamefont {J.~S.}\ \bibnamefont {Guasto}},
  \bibinfo {author} {\bibfnamefont {J.~P.}\ \bibnamefont {Gollub}}, \bibinfo
  {author} {\bibfnamefont {A.~I.}\ \bibnamefont {Pesci}}, \ and\ \bibinfo
  {author} {\bibfnamefont {R.~E.}\ \bibnamefont {Goldstein}},\ }\href {\doibase
  10.1103/PhysRevLett.103.198103} {\bibfield  {journal} {\bibinfo  {journal}
  {Physical Review Letters}\ }\textbf {\bibinfo {volume} {103}},\ \bibinfo
  {pages} {198103} (\bibinfo {year} {2009})}\BibitemShut {NoStop}%
\bibitem [{\citenamefont {Lagarde}\ \emph {et~al.}(2020)\citenamefont
  {Lagarde}, \citenamefont {Dag{\`e}s}, \citenamefont {Nemoto}, \citenamefont
  {D{\'e}mery}, \citenamefont {Bartolo},\ and\ \citenamefont
  {Gibaud}}]{lagarde2020colloidal}%
  \BibitemOpen
  \bibfield  {author} {\bibinfo {author} {\bibfnamefont {A.}~\bibnamefont
  {Lagarde}}, \bibinfo {author} {\bibfnamefont {N.}~\bibnamefont {Dag{\`e}s}},
  \bibinfo {author} {\bibfnamefont {T.}~\bibnamefont {Nemoto}}, \bibinfo
  {author} {\bibfnamefont {V.}~\bibnamefont {D{\'e}mery}}, \bibinfo {author}
  {\bibfnamefont {D.}~\bibnamefont {Bartolo}}, \ and\ \bibinfo {author}
  {\bibfnamefont {T.}~\bibnamefont {Gibaud}},\ }\href {\doibase
  10.1039/D0SM00309C} {\bibfield  {journal} {\bibinfo  {journal} {Soft Matter}\
  }\textbf {\bibinfo {volume} {16}},\ \bibinfo {pages} {7503} (\bibinfo {year}
  {2020})}\BibitemShut {NoStop}%
\bibitem [{\citenamefont {Humphries}\ \emph {et~al.}(2010)\citenamefont
  {Humphries}, \citenamefont {Queiroz}, \citenamefont {Dyer}, \citenamefont
  {Pade}, \citenamefont {Musyl}, \citenamefont {Schaefer}, \citenamefont
  {Fuller}, \citenamefont {Brunnschweiler}, \citenamefont {Doyle},
  \citenamefont {Houghton} \emph {et~al.}}]{humphries2010environmental}%
  \BibitemOpen
  \bibfield  {author} {\bibinfo {author} {\bibfnamefont {N.~E.}\ \bibnamefont
  {Humphries}}, \bibinfo {author} {\bibfnamefont {N.}~\bibnamefont {Queiroz}},
  \bibinfo {author} {\bibfnamefont {J.~R.}\ \bibnamefont {Dyer}}, \bibinfo
  {author} {\bibfnamefont {N.~G.}\ \bibnamefont {Pade}}, \bibinfo {author}
  {\bibfnamefont {M.~K.}\ \bibnamefont {Musyl}}, \bibinfo {author}
  {\bibfnamefont {K.~M.}\ \bibnamefont {Schaefer}}, \bibinfo {author}
  {\bibfnamefont {D.~W.}\ \bibnamefont {Fuller}}, \bibinfo {author}
  {\bibfnamefont {J.~M.}\ \bibnamefont {Brunnschweiler}}, \bibinfo {author}
  {\bibfnamefont {T.~K.}\ \bibnamefont {Doyle}}, \bibinfo {author}
  {\bibfnamefont {J.~D.}\ \bibnamefont {Houghton}},  \emph {et~al.},\ }\href
  {\doibase 10.1038/nature09116} {\bibfield  {journal} {\bibinfo  {journal}
  {Nature}\ }\textbf {\bibinfo {volume} {465}},\ \bibinfo {pages} {1066}
  (\bibinfo {year} {2010})}\BibitemShut {NoStop}%
\bibitem [{\citenamefont {Volpe}\ and\ \citenamefont
  {Volpe}(2017)}]{volpe2017topography}%
  \BibitemOpen
  \bibfield  {author} {\bibinfo {author} {\bibfnamefont {G.}~\bibnamefont
  {Volpe}}\ and\ \bibinfo {author} {\bibfnamefont {G.}~\bibnamefont {Volpe}},\
  }\href {https://doi.org/10.1073/pnas.1711371114} {\bibfield  {journal}
  {\bibinfo  {journal} {Proceedings of the National Academy of Sciences}\
  }\textbf {\bibinfo {volume} {114}},\ \bibinfo {pages} {11350} (\bibinfo
  {year} {2017})}\BibitemShut {NoStop}%
\bibitem [{\citenamefont {Shlesinger}\ and\ \citenamefont
  {Klafter}(1986)}]{shlesinger1986levy}%
  \BibitemOpen
  \bibfield  {author} {\bibinfo {author} {\bibfnamefont {M.~F.}\ \bibnamefont
  {Shlesinger}}\ and\ \bibinfo {author} {\bibfnamefont {J.}~\bibnamefont
  {Klafter}},\ }in\ \href {https://doi.org/10.1007/978-94-009-5165-5_29} {\emph
  {\bibinfo {booktitle} {On growth and form}}}\ (\bibinfo  {publisher}
  {Springer},\ \bibinfo {year} {1986})\ pp.\ \bibinfo {pages}
  {279--283}\BibitemShut {NoStop}%
\bibitem [{\citenamefont {Klafter}\ and\ \citenamefont
  {Sokolov}(2005)}]{klafter2005anomalous}%
  \BibitemOpen
  \bibfield  {author} {\bibinfo {author} {\bibfnamefont {J.}~\bibnamefont
  {Klafter}}\ and\ \bibinfo {author} {\bibfnamefont {I.~M.}\ \bibnamefont
  {Sokolov}},\ }\href {\doibase 10.1088/2058-7058/18/8/33} {\bibfield
  {journal} {\bibinfo  {journal} {Physics world}\ }\textbf {\bibinfo {volume}
  {18}},\ \bibinfo {pages} {29} (\bibinfo {year} {2005})}\BibitemShut {NoStop}%
\bibitem [{\citenamefont {Bartumeus}\ \emph {et~al.}(2002)\citenamefont
  {Bartumeus}, \citenamefont {Catalan}, \citenamefont {Fulco}, \citenamefont
  {Lyra},\ and\ \citenamefont {Viswanathan}}]{bartumeus2002optimizing}%
  \BibitemOpen
  \bibfield  {author} {\bibinfo {author} {\bibfnamefont {F.}~\bibnamefont
  {Bartumeus}}, \bibinfo {author} {\bibfnamefont {J.}~\bibnamefont {Catalan}},
  \bibinfo {author} {\bibfnamefont {U.}~\bibnamefont {Fulco}}, \bibinfo
  {author} {\bibfnamefont {M.}~\bibnamefont {Lyra}}, \ and\ \bibinfo {author}
  {\bibfnamefont {G.}~\bibnamefont {Viswanathan}},\ }\href {\doibase
  10.1103/PhysRevLett.88.097901} {\bibfield  {journal} {\bibinfo  {journal}
  {Physical Review Letters}\ }\textbf {\bibinfo {volume} {88}},\ \bibinfo
  {pages} {097901} (\bibinfo {year} {2002})}\BibitemShut {NoStop}%
\bibitem [{\citenamefont {Reynolds}(2015)}]{reynolds2015liberating}%
  \BibitemOpen
  \bibfield  {author} {\bibinfo {author} {\bibfnamefont {A.}~\bibnamefont
  {Reynolds}},\ }\href {\doibase 10.1016/j.plrev.2015.03.002} {\bibfield
  {journal} {\bibinfo  {journal} {Physics of life reviews}\ }\textbf {\bibinfo
  {volume} {14}},\ \bibinfo {pages} {59} (\bibinfo {year} {2015})}\BibitemShut
  {NoStop}%
\bibitem [{\citenamefont {Knebel}\ \emph {et~al.}(2021)\citenamefont {Knebel},
  \citenamefont {Sha-Ked}, \citenamefont {Agmon}, \citenamefont {Ariel},\ and\
  \citenamefont {Ayali}}]{knebel2021collective}%
  \BibitemOpen
  \bibfield  {author} {\bibinfo {author} {\bibfnamefont {D.}~\bibnamefont
  {Knebel}}, \bibinfo {author} {\bibfnamefont {C.}~\bibnamefont {Sha-Ked}},
  \bibinfo {author} {\bibfnamefont {N.}~\bibnamefont {Agmon}}, \bibinfo
  {author} {\bibfnamefont {G.}~\bibnamefont {Ariel}}, \ and\ \bibinfo {author}
  {\bibfnamefont {A.}~\bibnamefont {Ayali}},\ }\href
  {https://doi.org/10.1016/j.isci.2021.102299} {\bibfield  {journal} {\bibinfo
  {journal} {Iscience}\ }\textbf {\bibinfo {volume} {24}},\ \bibinfo {pages}
  {102299} (\bibinfo {year} {2021})}\BibitemShut {NoStop}%
\bibitem [{\citenamefont {Ariel}\ \emph {et~al.}(2015)\citenamefont {Ariel},
  \citenamefont {Rabani}, \citenamefont {Benisty}, \citenamefont {Partridge},
  \citenamefont {Harshey},\ and\ \citenamefont {Be'Er}}]{ariel2015swarming}%
  \BibitemOpen
  \bibfield  {author} {\bibinfo {author} {\bibfnamefont {G.}~\bibnamefont
  {Ariel}}, \bibinfo {author} {\bibfnamefont {A.}~\bibnamefont {Rabani}},
  \bibinfo {author} {\bibfnamefont {S.}~\bibnamefont {Benisty}}, \bibinfo
  {author} {\bibfnamefont {J.~D.}\ \bibnamefont {Partridge}}, \bibinfo {author}
  {\bibfnamefont {R.~M.}\ \bibnamefont {Harshey}}, \ and\ \bibinfo {author}
  {\bibfnamefont {A.}~\bibnamefont {Be'Er}},\ }\href {\doibase
  10.1038/ncomms9396} {\bibfield  {journal} {\bibinfo  {journal} {Nature
  communications}\ }\textbf {\bibinfo {volume} {6}},\ \bibinfo {pages} {1}
  (\bibinfo {year} {2015})}\BibitemShut {NoStop}%
\bibitem [{\citenamefont {Dombrowski}\ \emph {et~al.}(2004)\citenamefont
  {Dombrowski}, \citenamefont {Cisneros}, \citenamefont {Chatkaew},
  \citenamefont {Goldstein},\ and\ \citenamefont
  {Kessler}}]{dombrowski2004self}%
  \BibitemOpen
  \bibfield  {author} {\bibinfo {author} {\bibfnamefont {C.}~\bibnamefont
  {Dombrowski}}, \bibinfo {author} {\bibfnamefont {L.}~\bibnamefont
  {Cisneros}}, \bibinfo {author} {\bibfnamefont {S.}~\bibnamefont {Chatkaew}},
  \bibinfo {author} {\bibfnamefont {R.~E.}\ \bibnamefont {Goldstein}}, \ and\
  \bibinfo {author} {\bibfnamefont {J.~O.}\ \bibnamefont {Kessler}},\ }\href
  {\doibase 10.1103/PhysRevLett.93.098103} {\bibfield  {journal} {\bibinfo
  {journal} {Physical review letters}\ }\textbf {\bibinfo {volume} {93}},\
  \bibinfo {pages} {098103} (\bibinfo {year} {2004})}\BibitemShut {NoStop}%
\bibitem [{\citenamefont {Wensink}\ \emph {et~al.}(2012)\citenamefont
  {Wensink}, \citenamefont {Dunkel}, \citenamefont {Heidenreich}, \citenamefont
  {Drescher}, \citenamefont {Goldstein}, \citenamefont {L{\"o}wen},\ and\
  \citenamefont {Yeomans}}]{wensink2012meso}%
  \BibitemOpen
  \bibfield  {author} {\bibinfo {author} {\bibfnamefont {H.~H.}\ \bibnamefont
  {Wensink}}, \bibinfo {author} {\bibfnamefont {J.}~\bibnamefont {Dunkel}},
  \bibinfo {author} {\bibfnamefont {S.}~\bibnamefont {Heidenreich}}, \bibinfo
  {author} {\bibfnamefont {K.}~\bibnamefont {Drescher}}, \bibinfo {author}
  {\bibfnamefont {R.~E.}\ \bibnamefont {Goldstein}}, \bibinfo {author}
  {\bibfnamefont {H.}~\bibnamefont {L{\"o}wen}}, \ and\ \bibinfo {author}
  {\bibfnamefont {J.~M.}\ \bibnamefont {Yeomans}},\ }\href {\doibase
  10.1073/pnas.1202032109} {\bibfield  {journal} {\bibinfo  {journal}
  {Proceedings of the National Academy of Sciences}\ }\textbf {\bibinfo
  {volume} {109}},\ \bibinfo {pages} {14308} (\bibinfo {year}
  {2012})}\BibitemShut {NoStop}%
\bibitem [{\citenamefont {Dunkel}\ \emph
  {et~al.}(2013{\natexlab{a}})\citenamefont {Dunkel}, \citenamefont
  {Heidenreich}, \citenamefont {Drescher}, \citenamefont {Wensink},
  \citenamefont {B{\"a}r},\ and\ \citenamefont {Goldstein}}]{dunkel2013fluid}%
  \BibitemOpen
  \bibfield  {author} {\bibinfo {author} {\bibfnamefont {J.}~\bibnamefont
  {Dunkel}}, \bibinfo {author} {\bibfnamefont {S.}~\bibnamefont {Heidenreich}},
  \bibinfo {author} {\bibfnamefont {K.}~\bibnamefont {Drescher}}, \bibinfo
  {author} {\bibfnamefont {H.~H.}\ \bibnamefont {Wensink}}, \bibinfo {author}
  {\bibfnamefont {M.}~\bibnamefont {B{\"a}r}}, \ and\ \bibinfo {author}
  {\bibfnamefont {R.~E.}\ \bibnamefont {Goldstein}},\ }\href {\doibase
  10.1103/PhysRevLett.110.228102} {\bibfield  {journal} {\bibinfo  {journal}
  {Physical review letters}\ }\textbf {\bibinfo {volume} {110}},\ \bibinfo
  {pages} {228102} (\bibinfo {year} {2013}{\natexlab{a}})}\BibitemShut
  {NoStop}%
\bibitem [{\citenamefont {Zhou}\ \emph {et~al.}(2014)\citenamefont {Zhou},
  \citenamefont {Sokolov}, \citenamefont {Lavrentovich},\ and\ \citenamefont
  {Aranson}}]{zhou2014living}%
  \BibitemOpen
  \bibfield  {author} {\bibinfo {author} {\bibfnamefont {S.}~\bibnamefont
  {Zhou}}, \bibinfo {author} {\bibfnamefont {A.}~\bibnamefont {Sokolov}},
  \bibinfo {author} {\bibfnamefont {O.~D.}\ \bibnamefont {Lavrentovich}}, \
  and\ \bibinfo {author} {\bibfnamefont {I.~S.}\ \bibnamefont {Aranson}},\
  }\href {\doibase 10.1073/pnas.1321926111} {\bibfield  {journal} {\bibinfo
  {journal} {Proceedings of the National Academy of Sciences}\ }\textbf
  {\bibinfo {volume} {111}},\ \bibinfo {pages} {1265} (\bibinfo {year}
  {2014})}\BibitemShut {NoStop}%
\bibitem [{\citenamefont {Wu}\ \emph {et~al.}(2017)\citenamefont {Wu},
  \citenamefont {Hishamunda}, \citenamefont {Chen}, \citenamefont {DeCamp},
  \citenamefont {Chang}, \citenamefont {Fern{\'a}ndez-Nieves}, \citenamefont
  {Fraden},\ and\ \citenamefont {Dogic}}]{wu2017transition}%
  \BibitemOpen
  \bibfield  {author} {\bibinfo {author} {\bibfnamefont {K.-T.}\ \bibnamefont
  {Wu}}, \bibinfo {author} {\bibfnamefont {J.~B.}\ \bibnamefont {Hishamunda}},
  \bibinfo {author} {\bibfnamefont {D.~T.}\ \bibnamefont {Chen}}, \bibinfo
  {author} {\bibfnamefont {S.~J.}\ \bibnamefont {DeCamp}}, \bibinfo {author}
  {\bibfnamefont {Y.-W.}\ \bibnamefont {Chang}}, \bibinfo {author}
  {\bibfnamefont {A.}~\bibnamefont {Fern{\'a}ndez-Nieves}}, \bibinfo {author}
  {\bibfnamefont {S.}~\bibnamefont {Fraden}}, \ and\ \bibinfo {author}
  {\bibfnamefont {Z.}~\bibnamefont {Dogic}},\ }\href {\doibase
  10.1126/science.aal1979} {\bibfield  {journal} {\bibinfo  {journal}
  {Science}\ }\textbf {\bibinfo {volume} {355}} (\bibinfo {year} {2017}),\
  10.1126/science.aal1979}\BibitemShut {NoStop}%
\bibitem [{\citenamefont {Mart{\'\i}nez-Prat}\ \emph
  {et~al.}(2019)\citenamefont {Mart{\'\i}nez-Prat}, \citenamefont
  {Ign{\'e}s-Mullol}, \citenamefont {Casademunt},\ and\ \citenamefont
  {Sagu{\'e}s}}]{martinez2019selection}%
  \BibitemOpen
  \bibfield  {author} {\bibinfo {author} {\bibfnamefont {B.}~\bibnamefont
  {Mart{\'\i}nez-Prat}}, \bibinfo {author} {\bibfnamefont {J.}~\bibnamefont
  {Ign{\'e}s-Mullol}}, \bibinfo {author} {\bibfnamefont {J.}~\bibnamefont
  {Casademunt}}, \ and\ \bibinfo {author} {\bibfnamefont {F.}~\bibnamefont
  {Sagu{\'e}s}},\ }\href {\doibase 10.1038/s41567-018-0411-6} {\bibfield
  {journal} {\bibinfo  {journal} {Nature Physics}\ }\textbf {\bibinfo {volume}
  {15}},\ \bibinfo {pages} {362} (\bibinfo {year} {2019})}\BibitemShut
  {NoStop}%
\bibitem [{\citenamefont {Giomi}(2015)}]{giomi2015geometry}%
  \BibitemOpen
  \bibfield  {author} {\bibinfo {author} {\bibfnamefont {L.}~\bibnamefont
  {Giomi}},\ }\href {\doibase 10.1103/PhysRevX.5.031003} {\bibfield  {journal}
  {\bibinfo  {journal} {Physical Review X}\ }\textbf {\bibinfo {volume} {5}},\
  \bibinfo {pages} {031003} (\bibinfo {year} {2015})}\BibitemShut {NoStop}%
\bibitem [{\citenamefont {James}\ \emph {et~al.}(2018)\citenamefont {James},
  \citenamefont {Bos},\ and\ \citenamefont {Wilczek}}]{james2018turbulence}%
  \BibitemOpen
  \bibfield  {author} {\bibinfo {author} {\bibfnamefont {M.}~\bibnamefont
  {James}}, \bibinfo {author} {\bibfnamefont {W.~J.}\ \bibnamefont {Bos}}, \
  and\ \bibinfo {author} {\bibfnamefont {M.}~\bibnamefont {Wilczek}},\ }\href
  {\doibase 10.1103/PhysRevFluids.3.061101} {\bibfield  {journal} {\bibinfo
  {journal} {Physical Review Fluids}\ }\textbf {\bibinfo {volume} {3}},\
  \bibinfo {pages} {061101} (\bibinfo {year} {2018})}\BibitemShut {NoStop}%
\bibitem [{\citenamefont {James}\ and\ \citenamefont
  {Wilczek}(2018)}]{james2018vortex}%
  \BibitemOpen
  \bibfield  {author} {\bibinfo {author} {\bibfnamefont {M.}~\bibnamefont
  {James}}\ and\ \bibinfo {author} {\bibfnamefont {M.}~\bibnamefont
  {Wilczek}},\ }\href {\doibase 10.1140/epje/i2018-11625-8} {\bibfield
  {journal} {\bibinfo  {journal} {The European Physical Journal E}\ }\textbf
  {\bibinfo {volume} {41}},\ \bibinfo {pages} {1} (\bibinfo {year}
  {2018})}\BibitemShut {NoStop}%
\bibitem [{\citenamefont {Bratanov}\ \emph {et~al.}(2015)\citenamefont
  {Bratanov}, \citenamefont {Jenko},\ and\ \citenamefont
  {Frey}}]{bratanov2015new}%
  \BibitemOpen
  \bibfield  {author} {\bibinfo {author} {\bibfnamefont {V.}~\bibnamefont
  {Bratanov}}, \bibinfo {author} {\bibfnamefont {F.}~\bibnamefont {Jenko}}, \
  and\ \bibinfo {author} {\bibfnamefont {E.}~\bibnamefont {Frey}},\ }\href
  {\doibase 10.1073/pnas.1509304112} {\bibfield  {journal} {\bibinfo  {journal}
  {Proceedings of the National Academy of Sciences}\ }\textbf {\bibinfo
  {volume} {112}},\ \bibinfo {pages} {15048} (\bibinfo {year}
  {2015})}\BibitemShut {NoStop}%
\bibitem [{\citenamefont {Alert}\ \emph {et~al.}(2020)\citenamefont {Alert},
  \citenamefont {Joanny},\ and\ \citenamefont
  {Casademunt}}]{alert2020universal}%
  \BibitemOpen
  \bibfield  {author} {\bibinfo {author} {\bibfnamefont {R.}~\bibnamefont
  {Alert}}, \bibinfo {author} {\bibfnamefont {J.-F.}\ \bibnamefont {Joanny}}, \
  and\ \bibinfo {author} {\bibfnamefont {J.}~\bibnamefont {Casademunt}},\
  }\href {\doibase 10.1038/s41567-020-0854-4} {\bibfield  {journal} {\bibinfo
  {journal} {Nature Physics}\ }\textbf {\bibinfo {volume} {16}},\ \bibinfo
  {pages} {682} (\bibinfo {year} {2020})}\BibitemShut {NoStop}%
\bibitem [{\citenamefont {Alert}\ \emph {et~al.}(2021)\citenamefont {Alert},
  \citenamefont {Casademunt},\ and\ \citenamefont {Joanny}}]{alert2021}%
  \BibitemOpen
  \bibfield  {author} {\bibinfo {author} {\bibfnamefont {R.}~\bibnamefont
  {Alert}}, \bibinfo {author} {\bibfnamefont {J.}~\bibnamefont {Casademunt}}, \
  and\ \bibinfo {author} {\bibfnamefont {J.-F.}\ \bibnamefont {Joanny}},\
  }\href@noop {} {\bibfield  {journal} {\bibinfo  {journal} {arXiv preprint
  arXiv:2104.02122}\ } (\bibinfo {year} {2021})},\ \Eprint
  {http://arxiv.org/abs/2104.02122} {2104.02122} \BibitemShut {NoStop}%
\bibitem [{\citenamefont {Xia}\ \emph {et~al.}(2013)\citenamefont {Xia},
  \citenamefont {Francois}, \citenamefont {Punzmann},\ and\ \citenamefont
  {Shats}}]{xia2013lagrangian}%
  \BibitemOpen
  \bibfield  {author} {\bibinfo {author} {\bibfnamefont {H.}~\bibnamefont
  {Xia}}, \bibinfo {author} {\bibfnamefont {N.}~\bibnamefont {Francois}},
  \bibinfo {author} {\bibfnamefont {H.}~\bibnamefont {Punzmann}}, \ and\
  \bibinfo {author} {\bibfnamefont {M.}~\bibnamefont {Shats}},\ }\href
  {\doibase 10.1038/ncomms3013} {\bibfield  {journal} {\bibinfo  {journal}
  {Nature communications}\ }\textbf {\bibinfo {volume} {4}},\ \bibinfo {pages}
  {1} (\bibinfo {year} {2013})}\BibitemShut {NoStop}%
\bibitem [{\citenamefont {Wu}\ and\ \citenamefont
  {Libchaber}(2000)}]{wu2000particle}%
  \BibitemOpen
  \bibfield  {author} {\bibinfo {author} {\bibfnamefont {X.-L.}\ \bibnamefont
  {Wu}}\ and\ \bibinfo {author} {\bibfnamefont {A.}~\bibnamefont {Libchaber}},\
  }\href {\doibase 10.1103/PhysRevLett.84.3017} {\bibfield  {journal} {\bibinfo
   {journal} {Physical review letters}\ }\textbf {\bibinfo {volume} {84}},\
  \bibinfo {pages} {3017} (\bibinfo {year} {2000})}\BibitemShut {NoStop}%
\bibitem [{\citenamefont {Kurtuldu}\ \emph {et~al.}(2011)\citenamefont
  {Kurtuldu}, \citenamefont {Guasto}, \citenamefont {Johnson},\ and\
  \citenamefont {Gollub}}]{kurtuldu2011enhancement}%
  \BibitemOpen
  \bibfield  {author} {\bibinfo {author} {\bibfnamefont {H.}~\bibnamefont
  {Kurtuldu}}, \bibinfo {author} {\bibfnamefont {J.~S.}\ \bibnamefont
  {Guasto}}, \bibinfo {author} {\bibfnamefont {K.~A.}\ \bibnamefont {Johnson}},
  \ and\ \bibinfo {author} {\bibfnamefont {J.~P.}\ \bibnamefont {Gollub}},\
  }\href {\doibase 10.1073/pnas.1107046108} {\bibfield  {journal} {\bibinfo
  {journal} {Proceedings of the National Academy of Sciences}\ }\textbf
  {\bibinfo {volume} {108}},\ \bibinfo {pages} {10391} (\bibinfo {year}
  {2011})}\BibitemShut {NoStop}%
\bibitem [{\citenamefont {Morozov}\ and\ \citenamefont
  {Marenduzzo}(2014)}]{morozov2014enhanced}%
  \BibitemOpen
  \bibfield  {author} {\bibinfo {author} {\bibfnamefont {A.}~\bibnamefont
  {Morozov}}\ and\ \bibinfo {author} {\bibfnamefont {D.}~\bibnamefont
  {Marenduzzo}},\ }\href {\doibase 10.1039/C3SM52201F} {\bibfield  {journal}
  {\bibinfo  {journal} {Soft Matter}\ }\textbf {\bibinfo {volume} {10}},\
  \bibinfo {pages} {2748} (\bibinfo {year} {2014})}\BibitemShut {NoStop}%
\bibitem [{\citenamefont {Kanazawa}\ \emph {et~al.}(2020)\citenamefont
  {Kanazawa}, \citenamefont {Sano}, \citenamefont {Cairoli},\ and\
  \citenamefont {Baule}}]{kanazawa2020loopy}%
  \BibitemOpen
  \bibfield  {author} {\bibinfo {author} {\bibfnamefont {K.}~\bibnamefont
  {Kanazawa}}, \bibinfo {author} {\bibfnamefont {T.~G.}\ \bibnamefont {Sano}},
  \bibinfo {author} {\bibfnamefont {A.}~\bibnamefont {Cairoli}}, \ and\
  \bibinfo {author} {\bibfnamefont {A.}~\bibnamefont {Baule}},\ }\href
  {\doibase 10.1038/s41586-020-2086-2} {\bibfield  {journal} {\bibinfo
  {journal} {Nature}\ }\textbf {\bibinfo {volume} {579}},\ \bibinfo {pages}
  {364} (\bibinfo {year} {2020})}\BibitemShut {NoStop}%
\bibitem [{\citenamefont {Dunkel}\ \emph
  {et~al.}(2013{\natexlab{b}})\citenamefont {Dunkel}, \citenamefont
  {Heidenreich}, \citenamefont {B{\"a}r},\ and\ \citenamefont
  {Goldstein}}]{dunkel2013minimal}%
  \BibitemOpen
  \bibfield  {author} {\bibinfo {author} {\bibfnamefont {J.}~\bibnamefont
  {Dunkel}}, \bibinfo {author} {\bibfnamefont {S.}~\bibnamefont {Heidenreich}},
  \bibinfo {author} {\bibfnamefont {M.}~\bibnamefont {B{\"a}r}}, \ and\
  \bibinfo {author} {\bibfnamefont {R.~E.}\ \bibnamefont {Goldstein}},\ }\href
  {\doibase 10.1088/1367-2630/15/4/045016} {\bibfield  {journal} {\bibinfo
  {journal} {New Journal of Physics}\ }\textbf {\bibinfo {volume} {15}},\
  \bibinfo {pages} {045016} (\bibinfo {year} {2013}{\natexlab{b}})}\BibitemShut
  {NoStop}%
\bibitem [{\citenamefont {Swift}\ and\ \citenamefont
  {Hohenberg}(1977)}]{swift1977hydrodynamic}%
  \BibitemOpen
  \bibfield  {author} {\bibinfo {author} {\bibfnamefont {J.}~\bibnamefont
  {Swift}}\ and\ \bibinfo {author} {\bibfnamefont {P.~C.}\ \bibnamefont
  {Hohenberg}},\ }\href {\doibase 10.1103/PhysRevA.15.319} {\bibfield
  {journal} {\bibinfo  {journal} {Physical Review A}\ }\textbf {\bibinfo
  {volume} {15}},\ \bibinfo {pages} {319} (\bibinfo {year} {1977})}\BibitemShut
  {NoStop}%
\bibitem [{\citenamefont {Simha}\ and\ \citenamefont
  {Ramaswamy}(2002)}]{simha2002hydrodynamic}%
  \BibitemOpen
  \bibfield  {author} {\bibinfo {author} {\bibfnamefont {R.~A.}\ \bibnamefont
  {Simha}}\ and\ \bibinfo {author} {\bibfnamefont {S.}~\bibnamefont
  {Ramaswamy}},\ }\href {\doibase 10.1103/PhysRevLett.89.058101} {\bibfield
  {journal} {\bibinfo  {journal} {Physical review letters}\ }\textbf {\bibinfo
  {volume} {89}},\ \bibinfo {pages} {058101} (\bibinfo {year}
  {2002})}\BibitemShut {NoStop}%
\bibitem [{\citenamefont {Linkmann}\ \emph {et~al.}(2020)\citenamefont
  {Linkmann}, \citenamefont {Marchetti}, \citenamefont {Boffetta},\ and\
  \citenamefont {Eckhardt}}]{linkmann2020condensate}%
  \BibitemOpen
  \bibfield  {author} {\bibinfo {author} {\bibfnamefont {M.}~\bibnamefont
  {Linkmann}}, \bibinfo {author} {\bibfnamefont {M.~C.}\ \bibnamefont
  {Marchetti}}, \bibinfo {author} {\bibfnamefont {G.}~\bibnamefont {Boffetta}},
  \ and\ \bibinfo {author} {\bibfnamefont {B.}~\bibnamefont {Eckhardt}},\
  }\href {\doibase 10.1103/PhysRevE.101.022609} {\bibfield  {journal} {\bibinfo
   {journal} {Physical Review E}\ }\textbf {\bibinfo {volume} {101}},\ \bibinfo
  {pages} {022609} (\bibinfo {year} {2020})}\BibitemShut {NoStop}%
\bibitem [{\citenamefont {Toner}\ and\ \citenamefont {Tu}(1995)}]{TT95}%
  \BibitemOpen
  \bibfield  {author} {\bibinfo {author} {\bibfnamefont {J.}~\bibnamefont
  {Toner}}\ and\ \bibinfo {author} {\bibfnamefont {Y.}~\bibnamefont {Tu}},\
  }\href {\doibase 10.1103/PhysRevLett.75.4326} {\bibfield  {journal} {\bibinfo
   {journal} {Phys. Rev. Lett.}\ }\textbf {\bibinfo {volume} {75}},\ \bibinfo
  {pages} {4326} (\bibinfo {year} {1995})}\BibitemShut {NoStop}%
\bibitem [{\citenamefont {Toner}\ and\ \citenamefont {Tu}(1998)}]{TT98}%
  \BibitemOpen
  \bibfield  {author} {\bibinfo {author} {\bibfnamefont {J.}~\bibnamefont
  {Toner}}\ and\ \bibinfo {author} {\bibfnamefont {Y.}~\bibnamefont {Tu}},\
  }\href {\doibase 10.1103/PhysRevE.58.4828} {\bibfield  {journal} {\bibinfo
  {journal} {Phys. Rev. E}\ }\textbf {\bibinfo {volume} {58}},\ \bibinfo
  {pages} {4828} (\bibinfo {year} {1998})}\BibitemShut {NoStop}%
\bibitem [{\citenamefont {C.P.}\ and\ \citenamefont
  {Joy}(2020)}]{joy2020friction}%
  \BibitemOpen
  \bibfield  {author} {\bibinfo {author} {\bibfnamefont {S.}~\bibnamefont
  {C.P.}}\ and\ \bibinfo {author} {\bibfnamefont {A.}~\bibnamefont {Joy}},\
  }\href {\doibase 10.1103/PhysRevFluids.5.024302} {\bibfield  {journal}
  {\bibinfo  {journal} {Physical Review Fluids}\ }\textbf {\bibinfo {volume}
  {5}},\ \bibinfo {pages} {024302} (\bibinfo {year} {2020})}\BibitemShut
  {NoStop}%
\bibitem [{\citenamefont {Ariel}\ \emph {et~al.}(2017)\citenamefont {Ariel},
  \citenamefont {Be’er},\ and\ \citenamefont {Reynolds}}]{ariel2017chaotic}%
  \BibitemOpen
  \bibfield  {author} {\bibinfo {author} {\bibfnamefont {G.}~\bibnamefont
  {Ariel}}, \bibinfo {author} {\bibfnamefont {A.}~\bibnamefont {Be’er}}, \
  and\ \bibinfo {author} {\bibfnamefont {A.}~\bibnamefont {Reynolds}},\ }\href
  {https://doi.org/10.1103/PhysRevLett.118.228102} {\bibfield  {journal}
  {\bibinfo  {journal} {Physical review letters}\ }\textbf {\bibinfo {volume}
  {118}},\ \bibinfo {pages} {228102} (\bibinfo {year} {2017})}\BibitemShut
  {NoStop}%
\bibitem [{\citenamefont {Ariel}\ and\ \citenamefont
  {Schiff}(2020)}]{ariel2020conservative}%
  \BibitemOpen
  \bibfield  {author} {\bibinfo {author} {\bibfnamefont {G.}~\bibnamefont
  {Ariel}}\ and\ \bibinfo {author} {\bibfnamefont {J.}~\bibnamefont {Schiff}},\
  }\href {https://doi.org/10.1016/j.physd.2020.132584} {\bibfield  {journal}
  {\bibinfo  {journal} {Physica D: Nonlinear Phenomena}\ }\textbf {\bibinfo
  {volume} {411}},\ \bibinfo {pages} {132584} (\bibinfo {year}
  {2020})}\BibitemShut {NoStop}%
\bibitem [{\citenamefont {Panja}\ \emph {et~al.}(2015)\citenamefont {Panja},
  \citenamefont {Barkema},\ and\ \citenamefont {Ball}}]{panja2015complex}%
  \BibitemOpen
  \bibfield  {author} {\bibinfo {author} {\bibfnamefont {D.}~\bibnamefont
  {Panja}}, \bibinfo {author} {\bibfnamefont {G.~T.}\ \bibnamefont {Barkema}},
  \ and\ \bibinfo {author} {\bibfnamefont {R.~C.}\ \bibnamefont {Ball}},\
  }\href {https://doi.org/10.1021/ma502523p} {\bibfield  {journal} {\bibinfo
  {journal} {Macromolecules}\ }\textbf {\bibinfo {volume} {48}},\ \bibinfo
  {pages} {1442} (\bibinfo {year} {2015})}\BibitemShut {NoStop}%
\bibitem [{\citenamefont {van~der Hoeven}(2009)}]{hoeven}%
  \BibitemOpen
  \bibfield  {author} {\bibinfo {author} {\bibfnamefont {J.}~\bibnamefont
  {van~der Hoeven}},\ }\href@noop {} {\bibfield  {journal} {\bibinfo  {journal}
  {Journal of Symbolic Computation}\ }\textbf {\bibinfo {volume} {44}},\
  \bibinfo {pages} {1000} (\bibinfo {year} {2009})}\BibitemShut {NoStop}%
\bibitem [{\citenamefont {Pauls}\ and\ \citenamefont
  {Frisch}(2007)}]{pauls-frisch}%
  \BibitemOpen
  \bibfield  {author} {\bibinfo {author} {\bibfnamefont {W.}~\bibnamefont
  {Pauls}}\ and\ \bibinfo {author} {\bibfnamefont {U.}~\bibnamefont {Frisch}},\
  }\href {https://doi.org/10.1007/s10955-007-9307-z} {\bibfield  {journal}
  {\bibinfo  {journal} {J Stat Phys}\ }\textbf {\bibinfo {volume} {127}},\
  \bibinfo {pages} {1095} (\bibinfo {year} {2007})}\BibitemShut {NoStop}%
\bibitem [{\citenamefont {Bardos}\ \emph {et~al.}(2010)\citenamefont {Bardos},
  \citenamefont {Frisch}, \citenamefont {Pauls}, \citenamefont {Ray},\ and\
  \citenamefont {Titi}}]{entire}%
  \BibitemOpen
  \bibfield  {author} {\bibinfo {author} {\bibfnamefont {C.}~\bibnamefont
  {Bardos}}, \bibinfo {author} {\bibfnamefont {U.}~\bibnamefont {Frisch}},
  \bibinfo {author} {\bibfnamefont {W.}~\bibnamefont {Pauls}}, \bibinfo
  {author} {\bibfnamefont {S.~S.}\ \bibnamefont {Ray}}, \ and\ \bibinfo
  {author} {\bibfnamefont {E.}~\bibnamefont {Titi}},\ }\href
  {https://doi.org/10.1007/s00220-009-0916-z} {\bibfield  {journal} {\bibinfo
  {journal} {Commun. Math. Phys.}\ }\textbf {\bibinfo {volume} {293}},\
  \bibinfo {pages} {519} (\bibinfo {year} {2010})}\BibitemShut {NoStop}%
\bibitem [{\citenamefont {Chakraborty}\ \emph {et~al.}(2012)\citenamefont
  {Chakraborty}, \citenamefont {Frisch}, \citenamefont {Pauls},\ and\
  \citenamefont {Ray}}]{nelkin}%
  \BibitemOpen
  \bibfield  {author} {\bibinfo {author} {\bibfnamefont {S.}~\bibnamefont
  {Chakraborty}}, \bibinfo {author} {\bibfnamefont {U.}~\bibnamefont {Frisch}},
  \bibinfo {author} {\bibfnamefont {W.}~\bibnamefont {Pauls}}, \ and\ \bibinfo
  {author} {\bibfnamefont {S.~S.}\ \bibnamefont {Ray}},\ }\href
  {https://doi.org/10.1103/PhysRevE.85.015301} {\bibfield  {journal} {\bibinfo
  {journal} {Phys. Rev. E}\ }\textbf {\bibinfo {volume} {85}},\ \bibinfo
  {pages} {015301} (\bibinfo {year} {2012})}\BibitemShut {NoStop}%
\bibitem [{\citenamefont {Shlesinger}\ \emph {et~al.}(1987)\citenamefont
  {Shlesinger}, \citenamefont {West},\ and\ \citenamefont
  {Klafter}}]{shlesinger1987levy}%
  \BibitemOpen
  \bibfield  {author} {\bibinfo {author} {\bibfnamefont {M.~F.}\ \bibnamefont
  {Shlesinger}}, \bibinfo {author} {\bibfnamefont {B.}~\bibnamefont {West}}, \
  and\ \bibinfo {author} {\bibfnamefont {J.}~\bibnamefont {Klafter}},\ }\href
  {https://doi.org/10.1103/PhysRevLett.58.1100} {\bibfield  {journal} {\bibinfo
   {journal} {Physical Review Letters}\ }\textbf {\bibinfo {volume} {58}},\
  \bibinfo {pages} {1100} (\bibinfo {year} {1987})}\BibitemShut {NoStop}%
\bibitem [{\citenamefont {Zaburdaev}\ \emph {et~al.}(2015)\citenamefont
  {Zaburdaev}, \citenamefont {Denisov},\ and\ \citenamefont
  {Klafter}}]{zaburdaev2015levy}%
  \BibitemOpen
  \bibfield  {author} {\bibinfo {author} {\bibfnamefont {V.}~\bibnamefont
  {Zaburdaev}}, \bibinfo {author} {\bibfnamefont {S.}~\bibnamefont {Denisov}},
  \ and\ \bibinfo {author} {\bibfnamefont {J.}~\bibnamefont {Klafter}},\ }\href
  {\doibase 10.1103/RevModPhys.87.483} {\bibfield  {journal} {\bibinfo
  {journal} {Reviews of Modern Physics}\ }\textbf {\bibinfo {volume} {87}},\
  \bibinfo {pages} {483} (\bibinfo {year} {2015})}\BibitemShut {NoStop}%
\bibitem [{foo()}]{footnote1}%
  \BibitemOpen
  \href@noop {} {}\bibinfo {howpublished} {There are of course examples of more
  generalised L\'evy walks, such as those seen in the anomalous diffusion of
  cold atoms in optical lattices, where this linear relation
  fails~\cite{eli2012,eli2014}.}\BibitemShut {Stop}%
\bibitem [{\citenamefont {Pandit}\ \emph {et~al.}(2009)\citenamefont {Pandit},
  \citenamefont {Perlekar},\ and\ \citenamefont {Ray}}]{Pandit2009}%
  \BibitemOpen
  \bibfield  {author} {\bibinfo {author} {\bibfnamefont {R.}~\bibnamefont
  {Pandit}}, \bibinfo {author} {\bibfnamefont {P.}~\bibnamefont {Perlekar}}, \
  and\ \bibinfo {author} {\bibfnamefont {S.~S.}\ \bibnamefont {Ray}},\ }\href
  {\doibase 10.1007/s12043-009-0096-6} {\bibfield  {journal} {\bibinfo
  {journal} {Pramana}\ }\textbf {\bibinfo {volume} {73}},\ \bibinfo {pages}
  {157} (\bibinfo {year} {2009})}\BibitemShut {NoStop}%
\bibitem [{\citenamefont {Pandit}\ \emph {et~al.}(2017)\citenamefont {Pandit},
  \citenamefont {Banerjee}, \citenamefont {Bhatnagar}, \citenamefont {Brachet},
  \citenamefont {Gupta}, \citenamefont {Mitra}, \citenamefont {Pal},
  \citenamefont {Perlekar}, \citenamefont {Ray}, \citenamefont {Shukla} \emph
  {et~al.}}]{pandit2017overview}%
  \BibitemOpen
  \bibfield  {author} {\bibinfo {author} {\bibfnamefont {R.}~\bibnamefont
  {Pandit}}, \bibinfo {author} {\bibfnamefont {D.}~\bibnamefont {Banerjee}},
  \bibinfo {author} {\bibfnamefont {A.}~\bibnamefont {Bhatnagar}}, \bibinfo
  {author} {\bibfnamefont {M.}~\bibnamefont {Brachet}}, \bibinfo {author}
  {\bibfnamefont {A.}~\bibnamefont {Gupta}}, \bibinfo {author} {\bibfnamefont
  {D.}~\bibnamefont {Mitra}}, \bibinfo {author} {\bibfnamefont
  {N.}~\bibnamefont {Pal}}, \bibinfo {author} {\bibfnamefont {P.}~\bibnamefont
  {Perlekar}}, \bibinfo {author} {\bibfnamefont {S.~S.}\ \bibnamefont {Ray}},
  \bibinfo {author} {\bibfnamefont {V.}~\bibnamefont {Shukla}},  \emph
  {et~al.},\ }\href {https://doi.org/10.1063/1.4986802} {\bibfield  {journal}
  {\bibinfo  {journal} {Physics of fluids}\ }\textbf {\bibinfo {volume} {29}},\
  \bibinfo {pages} {111112} (\bibinfo {year} {2017})}\BibitemShut {NoStop}%
\bibitem [{\citenamefont {Wensink}\ and\ \citenamefont
  {L{\"o}wen}(2012)}]{wensink2012emergent}%
  \BibitemOpen
  \bibfield  {author} {\bibinfo {author} {\bibfnamefont {H.}~\bibnamefont
  {Wensink}}\ and\ \bibinfo {author} {\bibfnamefont {H.}~\bibnamefont
  {L{\"o}wen}},\ }\href {https://doi.org/10.1088/0953-8984/24/46/464130}
  {\bibfield  {journal} {\bibinfo  {journal} {Journal of Physics: Condensed
  Matter}\ }\textbf {\bibinfo {volume} {24}},\ \bibinfo {pages} {464130}
  (\bibinfo {year} {2012})}\BibitemShut {NoStop}%
\bibitem [{\citenamefont {Chen}\ \emph {et~al.}(2016)\citenamefont {Chen},
  \citenamefont {Lee},\ and\ \citenamefont {Toner}}]{toner}%
  \BibitemOpen
  \bibfield  {author} {\bibinfo {author} {\bibfnamefont {L.}~\bibnamefont
  {Chen}}, \bibinfo {author} {\bibfnamefont {C.~F.}\ \bibnamefont {Lee}}, \
  and\ \bibinfo {author} {\bibfnamefont {J.}~\bibnamefont {Toner}},\ }\href
  {https://www.nature.com/articles/ncomms12215#citeas} {\bibfield  {journal}
  {\bibinfo  {journal} {Nature communications}\ }\textbf {\bibinfo {volume}
  {7}},\ \bibinfo {pages} {12215} (\bibinfo {year} {2016})}\BibitemShut
  {NoStop}%
\bibitem [{\citenamefont {Maitra}(2021)}]{PC-Maitra}%
  \BibitemOpen
  \bibfield  {author} {\bibinfo {author} {\bibfnamefont {A.}~\bibnamefont
  {Maitra}},\ }\href@noop {} {}\bibinfo {howpublished} {{Private
  Communication}} (\bibinfo {year} {2021})\BibitemShut {NoStop}%
\bibitem [{\citenamefont {Reynolds}(2018)}]{reynolds2018current}%
  \BibitemOpen
  \bibfield  {author} {\bibinfo {author} {\bibfnamefont {A.~M.}\ \bibnamefont
  {Reynolds}},\ }\href {\doibase 10.1242/bio.030106} {\bibfield  {journal}
  {\bibinfo  {journal} {Biology open}\ }\textbf {\bibinfo {volume} {7}}
  (\bibinfo {year} {2018}),\ 10.1242/bio.030106}\BibitemShut {NoStop}%
\bibitem [{\citenamefont {Reynolds}\ and\ \citenamefont
  {Ouellette}(2016)}]{reynolds2016swarm}%
  \BibitemOpen
  \bibfield  {author} {\bibinfo {author} {\bibfnamefont {A.~M.}\ \bibnamefont
  {Reynolds}}\ and\ \bibinfo {author} {\bibfnamefont {N.~T.}\ \bibnamefont
  {Ouellette}},\ }\href {\doibase 10.1038/srep30515} {\bibfield  {journal}
  {\bibinfo  {journal} {Scientific reports}\ }\textbf {\bibinfo {volume} {6}},\
  \bibinfo {pages} {1} (\bibinfo {year} {2016})}\BibitemShut {NoStop}%
\bibitem [{\citenamefont {Huda}\ \emph {et~al.}(2018)\citenamefont {Huda},
  \citenamefont {Weigelin}, \citenamefont {Wolf}, \citenamefont {Tretiakov},
  \citenamefont {Polev}, \citenamefont {Wilk}, \citenamefont {Iwasa},
  \citenamefont {Emami}, \citenamefont {Narojczyk}, \citenamefont {Banaszak}
  \emph {et~al.}}]{huda2018levy}%
  \BibitemOpen
  \bibfield  {author} {\bibinfo {author} {\bibfnamefont {S.}~\bibnamefont
  {Huda}}, \bibinfo {author} {\bibfnamefont {B.}~\bibnamefont {Weigelin}},
  \bibinfo {author} {\bibfnamefont {K.}~\bibnamefont {Wolf}}, \bibinfo {author}
  {\bibfnamefont {K.~V.}\ \bibnamefont {Tretiakov}}, \bibinfo {author}
  {\bibfnamefont {K.}~\bibnamefont {Polev}}, \bibinfo {author} {\bibfnamefont
  {G.}~\bibnamefont {Wilk}}, \bibinfo {author} {\bibfnamefont {M.}~\bibnamefont
  {Iwasa}}, \bibinfo {author} {\bibfnamefont {F.~S.}\ \bibnamefont {Emami}},
  \bibinfo {author} {\bibfnamefont {J.~W.}\ \bibnamefont {Narojczyk}}, \bibinfo
  {author} {\bibfnamefont {M.}~\bibnamefont {Banaszak}},  \emph {et~al.},\
  }\href {\doibase 10.1038/s41467-018-06563-w} {\bibfield  {journal} {\bibinfo
  {journal} {Nature communications}\ }\textbf {\bibinfo {volume} {9}},\
  \bibinfo {pages} {1} (\bibinfo {year} {2018})}\BibitemShut {NoStop}%
\bibitem [{\citenamefont {Gal}\ and\ \citenamefont {Weihs}(2010)}]{gal2010}%
  \BibitemOpen
  \bibfield  {author} {\bibinfo {author} {\bibfnamefont {N.}~\bibnamefont
  {Gal}}\ and\ \bibinfo {author} {\bibfnamefont {D.}~\bibnamefont {Weihs}},\
  }\href {\doibase 10.1103/PhysRevE.81.020903} {\bibfield  {journal} {\bibinfo
  {journal} {Phys. Rev. E}\ }\textbf {\bibinfo {volume} {81}},\ \bibinfo
  {pages} {020903} (\bibinfo {year} {2010})}\BibitemShut {NoStop}%
\bibitem [{\citenamefont {Muralidharan}\ \emph {et~al.}(2021)\citenamefont
  {Muralidharan}, \citenamefont {Uitenbroek},\ and\ \citenamefont
  {Boukany}}]{muralidharan2021intracellular}%
  \BibitemOpen
  \bibfield  {author} {\bibinfo {author} {\bibfnamefont {A.}~\bibnamefont
  {Muralidharan}}, \bibinfo {author} {\bibfnamefont {H.}~\bibnamefont
  {Uitenbroek}}, \ and\ \bibinfo {author} {\bibfnamefont {P.~E.}\ \bibnamefont
  {Boukany}},\ }\href {\doibase 10.1101/2021.04.12.435513} {\bibfield
  {journal} {\bibinfo  {journal} {bioRxiv}\ } (\bibinfo {year} {2021}),\
  10.1101/2021.04.12.435513}\BibitemShut {NoStop}%
\bibitem [{\citenamefont {Sims}\ \emph {et~al.}(2008)\citenamefont {Sims},
  \citenamefont {Southall}, \citenamefont {Humphries}, \citenamefont {Hays},
  \citenamefont {Bradshaw}, \citenamefont {Pitchford}, \citenamefont {James},
  \citenamefont {Ahmed}, \citenamefont {Brierley}, \citenamefont {Hindell}
  \emph {et~al.}}]{sims2008scaling}%
  \BibitemOpen
  \bibfield  {author} {\bibinfo {author} {\bibfnamefont {D.~W.}\ \bibnamefont
  {Sims}}, \bibinfo {author} {\bibfnamefont {E.~J.}\ \bibnamefont {Southall}},
  \bibinfo {author} {\bibfnamefont {N.~E.}\ \bibnamefont {Humphries}}, \bibinfo
  {author} {\bibfnamefont {G.~C.}\ \bibnamefont {Hays}}, \bibinfo {author}
  {\bibfnamefont {C.~J.}\ \bibnamefont {Bradshaw}}, \bibinfo {author}
  {\bibfnamefont {J.~W.}\ \bibnamefont {Pitchford}}, \bibinfo {author}
  {\bibfnamefont {A.}~\bibnamefont {James}}, \bibinfo {author} {\bibfnamefont
  {M.~Z.}\ \bibnamefont {Ahmed}}, \bibinfo {author} {\bibfnamefont {A.~S.}\
  \bibnamefont {Brierley}}, \bibinfo {author} {\bibfnamefont {M.~A.}\
  \bibnamefont {Hindell}},  \emph {et~al.},\ }\href {\doibase
  10.1038/nature06518} {\bibfield  {journal} {\bibinfo  {journal} {Nature}\
  }\textbf {\bibinfo {volume} {451}},\ \bibinfo {pages} {1098} (\bibinfo {year}
  {2008})}\BibitemShut {NoStop}%
\bibitem [{\citenamefont {Reijers}\ \emph {et~al.}(2019)\citenamefont
  {Reijers}, \citenamefont {Siteur}, \citenamefont {Hoeks}, \citenamefont {van
  Belzen}, \citenamefont {Borst}, \citenamefont {Heusinkveld}, \citenamefont
  {Govers}, \citenamefont {Bouma}, \citenamefont {Lamers}, \citenamefont
  {van~de Koppel} \emph {et~al.}}]{reijers2019levy}%
  \BibitemOpen
  \bibfield  {author} {\bibinfo {author} {\bibfnamefont {V.~C.}\ \bibnamefont
  {Reijers}}, \bibinfo {author} {\bibfnamefont {K.}~\bibnamefont {Siteur}},
  \bibinfo {author} {\bibfnamefont {S.}~\bibnamefont {Hoeks}}, \bibinfo
  {author} {\bibfnamefont {J.}~\bibnamefont {van Belzen}}, \bibinfo {author}
  {\bibfnamefont {A.~C.}\ \bibnamefont {Borst}}, \bibinfo {author}
  {\bibfnamefont {J.~H.}\ \bibnamefont {Heusinkveld}}, \bibinfo {author}
  {\bibfnamefont {L.~L.}\ \bibnamefont {Govers}}, \bibinfo {author}
  {\bibfnamefont {T.~J.}\ \bibnamefont {Bouma}}, \bibinfo {author}
  {\bibfnamefont {L.~P.}\ \bibnamefont {Lamers}}, \bibinfo {author}
  {\bibfnamefont {J.}~\bibnamefont {van~de Koppel}},  \emph {et~al.},\ }\href
  {\doibase 10.1038/s41467-019-10699-8} {\bibfield  {journal} {\bibinfo
  {journal} {Nature communications}\ }\textbf {\bibinfo {volume} {10}},\
  \bibinfo {pages} {1} (\bibinfo {year} {2019})}\BibitemShut {NoStop}%
\bibitem [{\citenamefont {Heidenreich}\ \emph {et~al.}(2016)\citenamefont
  {Heidenreich}, \citenamefont {Dunkel}, \citenamefont {Klapp},\ and\
  \citenamefont {B{\"a}r}}]{heidenreich2016hydrodynamic}%
  \BibitemOpen
  \bibfield  {author} {\bibinfo {author} {\bibfnamefont {S.}~\bibnamefont
  {Heidenreich}}, \bibinfo {author} {\bibfnamefont {J.}~\bibnamefont {Dunkel}},
  \bibinfo {author} {\bibfnamefont {S.~H.}\ \bibnamefont {Klapp}}, \ and\
  \bibinfo {author} {\bibfnamefont {M.}~\bibnamefont {B{\"a}r}},\ }\href
  {\doibase 10.1103/PhysRevE.94.020601} {\bibfield  {journal} {\bibinfo
  {journal} {Physical Review E}\ }\textbf {\bibinfo {volume} {94}},\ \bibinfo
  {pages} {020601} (\bibinfo {year} {2016})}\BibitemShut {NoStop}%
\bibitem [{\citenamefont {Reinken}\ \emph {et~al.}(2018)\citenamefont
  {Reinken}, \citenamefont {Klapp}, \citenamefont {B{\"a}r},\ and\
  \citenamefont {Heidenreich}}]{reinken2018derivation}%
  \BibitemOpen
  \bibfield  {author} {\bibinfo {author} {\bibfnamefont {H.}~\bibnamefont
  {Reinken}}, \bibinfo {author} {\bibfnamefont {S.~H.}\ \bibnamefont {Klapp}},
  \bibinfo {author} {\bibfnamefont {M.}~\bibnamefont {B{\"a}r}}, \ and\
  \bibinfo {author} {\bibfnamefont {S.}~\bibnamefont {Heidenreich}},\ }\href
  {\doibase 10.1103/PhysRevE.97.022613} {\bibfield  {journal} {\bibinfo
  {journal} {Physical Review E}\ }\textbf {\bibinfo {volume} {97}},\ \bibinfo
  {pages} {022613} (\bibinfo {year} {2018})}\BibitemShut {NoStop}%
\bibitem [{\citenamefont {James}\ \emph {et~al.}(2020)\citenamefont {James},
  \citenamefont {Suchla}, \citenamefont {Dunkel},\ and\ \citenamefont
  {Wilczek}}]{james2020emergence}%
  \BibitemOpen
  \bibfield  {author} {\bibinfo {author} {\bibfnamefont {M.}~\bibnamefont
  {James}}, \bibinfo {author} {\bibfnamefont {D.~A.}\ \bibnamefont {Suchla}},
  \bibinfo {author} {\bibfnamefont {J.}~\bibnamefont {Dunkel}}, \ and\ \bibinfo
  {author} {\bibfnamefont {M.}~\bibnamefont {Wilczek}},\ }\href@noop {}
  {\bibfield  {journal} {\bibinfo  {journal} {arXiv preprint arXiv:2005.06217}\
  } (\bibinfo {year} {2020})},\ \Eprint {http://arxiv.org/abs/2005.06217}
  {2005.06217} \BibitemShut {NoStop}%
\bibitem [{\citenamefont {Creppy}\ \emph {et~al.}(2015)\citenamefont {Creppy},
  \citenamefont {Praud}, \citenamefont {Druart}, \citenamefont {Kohnke},\ and\
  \citenamefont {Plourabou{\'e}}}]{creppy2015turbulence}%
  \BibitemOpen
  \bibfield  {author} {\bibinfo {author} {\bibfnamefont {A.}~\bibnamefont
  {Creppy}}, \bibinfo {author} {\bibfnamefont {O.}~\bibnamefont {Praud}},
  \bibinfo {author} {\bibfnamefont {X.}~\bibnamefont {Druart}}, \bibinfo
  {author} {\bibfnamefont {P.~L.}\ \bibnamefont {Kohnke}}, \ and\ \bibinfo
  {author} {\bibfnamefont {F.}~\bibnamefont {Plourabou{\'e}}},\ }\href
  {\doibase 10.1103/PhysRevE.92.032722} {\bibfield  {journal} {\bibinfo
  {journal} {Physical Review E}\ }\textbf {\bibinfo {volume} {92}},\ \bibinfo
  {pages} {032722} (\bibinfo {year} {2015})}\BibitemShut {NoStop}%
\bibitem [{\citenamefont {Alavi}\ and\ \citenamefont
  {Cosson}(2005)}]{alavi2005sperm}%
  \BibitemOpen
  \bibfield  {author} {\bibinfo {author} {\bibfnamefont {S.~M.~H.}\
  \bibnamefont {Alavi}}\ and\ \bibinfo {author} {\bibfnamefont
  {J.}~\bibnamefont {Cosson}},\ }\href {\doibase
  https://doi.org/10.1016/j.cellbi.2004.11.021} {\bibfield  {journal} {\bibinfo
   {journal} {Cell biology international}\ }\textbf {\bibinfo {volume} {29}},\
  \bibinfo {pages} {101} (\bibinfo {year} {2005})}\BibitemShut {NoStop}%
\bibitem [{\citenamefont {Boffetta}\ and\ \citenamefont
  {Celani}(2000)}]{boffetta2000pair}%
  \BibitemOpen
  \bibfield  {author} {\bibinfo {author} {\bibfnamefont {G.}~\bibnamefont
  {Boffetta}}\ and\ \bibinfo {author} {\bibfnamefont {A.}~\bibnamefont
  {Celani}},\ }\href {https://doi.org/10.1016/S0378-4371(99)00613-5} {\bibfield
   {journal} {\bibinfo  {journal} {Physica A: Statistical Mechanics and its
  Applications}\ }\textbf {\bibinfo {volume} {280}},\ \bibinfo {pages} {1}
  (\bibinfo {year} {2000})}\BibitemShut {NoStop}%
\bibitem [{\citenamefont {Spohn}(2014)}]{spohn}%
  \BibitemOpen
  \bibfield  {author} {\bibinfo {author} {\bibfnamefont {H.}~\bibnamefont
  {Spohn}},\ }\href {\doibase 10.1007/s10955-014-0933-y} {\bibfield  {journal}
  {\bibinfo  {journal} {J Stat Phys}\ }\textbf {\bibinfo {volume} {154}},\
  \bibinfo {pages} {1191} (\bibinfo {year} {2014})}\BibitemShut {NoStop}%
\bibitem [{\citenamefont {Miron}(2020{\natexlab{a}})}]{MironPRL2020}%
  \BibitemOpen
  \bibfield  {author} {\bibinfo {author} {\bibfnamefont {A.}~\bibnamefont
  {Miron}},\ }\href {\doibase 10.1103/PhysRevLett.124.140601} {\bibfield
  {journal} {\bibinfo  {journal} {Phys. Rev. Lett.}\ }\textbf {\bibinfo
  {volume} {124}},\ \bibinfo {pages} {140601} (\bibinfo {year}
  {2020}{\natexlab{a}})}\BibitemShut {NoStop}%
\bibitem [{\citenamefont {Miron}(2020{\natexlab{b}})}]{MironPRR2020}%
  \BibitemOpen
  \bibfield  {author} {\bibinfo {author} {\bibfnamefont {A.}~\bibnamefont
  {Miron}},\ }\href {\doibase 10.1103/PhysRevResearch.2.032042} {\bibfield
  {journal} {\bibinfo  {journal} {Phys. Rev. Research}\ }\textbf {\bibinfo
  {volume} {2}},\ \bibinfo {pages} {032042} (\bibinfo {year}
  {2020}{\natexlab{b}})}\BibitemShut {NoStop}%
\bibitem [{\citenamefont {Miron}(2021)}]{PC-Miron}%
  \BibitemOpen
  \bibfield  {author} {\bibinfo {author} {\bibfnamefont {A.}~\bibnamefont
  {Miron}},\ }\href@noop {} {}\bibinfo {howpublished} {{Private Communication}}
  (\bibinfo {year} {2021})\BibitemShut {NoStop}%
\bibitem [{\citenamefont {Perlekar}\ \emph {et~al.}(2011)\citenamefont
  {Perlekar}, \citenamefont {Ray}, \citenamefont {Mitra},\ and\ \citenamefont
  {Pandit}}]{perl2011}%
  \BibitemOpen
  \bibfield  {author} {\bibinfo {author} {\bibfnamefont {P.}~\bibnamefont
  {Perlekar}}, \bibinfo {author} {\bibfnamefont {S.~S.}\ \bibnamefont {Ray}},
  \bibinfo {author} {\bibfnamefont {D.}~\bibnamefont {Mitra}}, \ and\ \bibinfo
  {author} {\bibfnamefont {R.}~\bibnamefont {Pandit}},\ }\href
  {https://doi.org/10.1103/PhysRevLett.106.054501} {\bibfield  {journal}
  {\bibinfo  {journal} {Physical review letters}\ }\textbf {\bibinfo {volume}
  {106}},\ \bibinfo {pages} {054501} (\bibinfo {year} {2011})}\BibitemShut
  {NoStop}%
\bibitem [{\citenamefont {Ray}\ \emph {et~al.}(2011)\citenamefont {Ray},
  \citenamefont {Mitra}, \citenamefont {Perlekar},\ and\ \citenamefont
  {Pandit}}]{ssr2011}%
  \BibitemOpen
  \bibfield  {author} {\bibinfo {author} {\bibfnamefont {S.~S.}\ \bibnamefont
  {Ray}}, \bibinfo {author} {\bibfnamefont {D.}~\bibnamefont {Mitra}}, \bibinfo
  {author} {\bibfnamefont {P.}~\bibnamefont {Perlekar}}, \ and\ \bibinfo
  {author} {\bibfnamefont {R.}~\bibnamefont {Pandit}},\ }\href
  {https://link.aps.org/doi/10.1103/PhysRevLett.107.184503} {\bibfield
  {journal} {\bibinfo  {journal} {Phys. Rev. Lett.}\ }\textbf {\bibinfo
  {volume} {107}},\ \bibinfo {pages} {184503} (\bibinfo {year}
  {2011})}\BibitemShut {NoStop}%
\bibitem [{\citenamefont {Kessler}\ and\ \citenamefont
  {Barkai}(2012)}]{eli2012}%
  \BibitemOpen
  \bibfield  {author} {\bibinfo {author} {\bibfnamefont {D.~A.}\ \bibnamefont
  {Kessler}}\ and\ \bibinfo {author} {\bibfnamefont {E.}~\bibnamefont
  {Barkai}},\ }\href {\doibase 10.1103/PhysRevLett.108.230602} {\bibfield
  {journal} {\bibinfo  {journal} {Phys. Rev. Lett.}\ }\textbf {\bibinfo
  {volume} {108}},\ \bibinfo {pages} {230602} (\bibinfo {year}
  {2012})}\BibitemShut {NoStop}%
\bibitem [{\citenamefont {Barkai}\ \emph {et~al.}(2014)\citenamefont {Barkai},
  \citenamefont {Aghion},\ and\ \citenamefont {Kessler}}]{eli2014}%
  \BibitemOpen
  \bibfield  {author} {\bibinfo {author} {\bibfnamefont {E.}~\bibnamefont
  {Barkai}}, \bibinfo {author} {\bibfnamefont {E.}~\bibnamefont {Aghion}}, \
  and\ \bibinfo {author} {\bibfnamefont {D.~A.}\ \bibnamefont {Kessler}},\
  }\href {\doibase 10.1103/PhysRevX.4.021036} {\bibfield  {journal} {\bibinfo
  {journal} {Phys. Rev. X}\ }\textbf {\bibinfo {volume} {4}},\ \bibinfo {pages}
  {021036} (\bibinfo {year} {2014})}\BibitemShut {NoStop}%
\end{thebibliography}%

\end{document}